\documentclass[aps,prd,twocolumn,superscriptaddress,showpacs,bibnotes]{revtex4}
\usepackage{graphicx}
\usepackage{amsmath}

\begin{document}

\preprint{IceCube draft version 4.0}

\title{Measurement of the atmospheric neutrino energy spectrum from 100 GeV to 400 TeV with IceCube}

\affiliation{III. Physikalisches Institut, RWTH Aachen University, D-52056 Aachen, Germany}
\affiliation{Department of Physics and Astronomy, University of Alabama, Tuscaloosa, Alabama 35487, USA}
\affiliation{Department of Physics and Astronomy, University of Alaska Anchorage, 3211 Providence Dr., Anchorage, Alaska 99508, USA}
\affiliation{CTSPS, Clark-Atlanta University, Atlanta, Georgia 30314, USA}
\affiliation{School of Physics and Center for Relativistic Astrophysics, Georgia Institute of Technology, Atlanta, Georgia 30332, USA}
\affiliation{Department of Physics, Southern University, Baton Rouge, Louisiana 70813, USA}
\affiliation{Department of Physics, University of California, Berkeley, California 94720, USA}
\affiliation{Lawrence Berkeley National Laboratory, Berkeley, California 94720, USA}
\affiliation{Institut f\"ur Physik, Humboldt-Universit\"at zu Berlin, D-12489 Berlin, Germany}
\affiliation{Fakult\"at f\"ur Physik \& Astronomie, Ruhr-Universit\"at Bochum, D-44780 Bochum, Germany}
\affiliation{Physikalisches Institut, Universit\"at Bonn, Nussallee 12, D-53115 Bonn, Germany}
\affiliation{Department of Physics, University of the West Indies, Cave Hill Campus, Bridgetown BB11000, Barbados}
\affiliation{Universit\'e Libre de Bruxelles, Science Faculty CP230, B-1050 Brussels, Belgium}
\affiliation{Vrije Universiteit Brussel, Dienst ELEM, B-1050 Brussels, Belgium}
\affiliation{Department of Physics, Chiba University, Chiba 263-8522, Japan}
\affiliation{Department of Physics and Astronomy, University of Canterbury, Private Bag 4800, Christchurch, New Zealand}
\affiliation{Department of Physics, University of Maryland, College Park, Maryland 20742, USA}
\affiliation{Department of Physics and Center for Cosmology and Astro-Particle Physics, Ohio State University, Columbus, Ohio 43210, USA}
\affiliation{Department of Astronomy, Ohio State University, Columbus, Ohio 43210, USA}
\affiliation{Department of Physics, TU Dortmund University, D-44221 Dortmund, Germany}
\affiliation{Department of Physics, University of Alberta, Edmonton, Alberta, Canada T6G 2G7}
\affiliation{Department of Subatomic and Radiation Physics, University of Gent, B-9000 Gent, Belgium}
\affiliation{Max-Planck-Institut f\"ur Kernphysik, D-69177 Heidelberg, Germany}
\affiliation{Department of Physics and Astronomy, University of California, Irvine, California 92697, USA}
\affiliation{Laboratory for High Energy Physics, \'Ecole Polytechnique F\'ed\'erale, CH-1015 Lausanne, Switzerland}
\affiliation{Department of Physics and Astronomy, University of Kansas, Lawrence, Kansas 66045, USA}
\affiliation{Department of Astronomy, University of Wisconsin, Madison, Wisconsin 53706, USA}
\affiliation{Department of Physics, University of Wisconsin, Madison, Wisconsin 53706, USA}
\affiliation{Institute of Physics, University of Mainz, Staudinger Weg 7, D-55099 Mainz, Germany}
\affiliation{Universit\'e de Mons, 7000 Mons, Belgium}
\affiliation{Bartol Research Institute and Department of Physics and Astronomy, University of Delaware, Newark, Delaware 19716, USA}
\affiliation{Department of Physics, University of Oxford, 1 Keble Road, Oxford OX1 3NP, UK}
\affiliation{Department of Physics, University of Wisconsin, River Falls, Wisconsin 54022, USA}
\affiliation{Oskar Klein Centre and Department of Physics, Stockholm University, SE-10691 Stockholm, Sweden}
\affiliation{Department of Astronomy and Astrophysics, Pennsylvania State University, University Park, Pennsylvania 16802, USA}
\affiliation{Department of Physics, Pennsylvania State University, University Park, Pennsylvania 16802, USA}
\affiliation{Department of Physics and Astronomy, Uppsala University, Box 516, S-75120 Uppsala, Sweden}
\affiliation{Department of Physics and Astronomy, Utrecht University/SRON, NL-3584 CC Utrecht, The Netherlands}
\affiliation{Department of Physics, University of Wuppertal, D-42119 Wuppertal, Germany}
\affiliation{DESY, D-15735 Zeuthen, Germany}

\author{R.~Abbasi}
\affiliation{Department of Physics, University of Wisconsin, Madison, Wisconsin 53706, USA}
\author{Y.~Abdou}
\affiliation{Department of Subatomic and Radiation Physics, University of Gent, B-9000 Gent, Belgium}
\author{T.~Abu-Zayyad}
\affiliation{Department of Physics, University of Wisconsin, River Falls, Wisconsin 54022, USA}
\author{J.~Adams}
\affiliation{Department of Physics and Astronomy, University of Canterbury, Private Bag 4800, Christchurch, New Zealand}
\author{J.~A.~Aguilar}
\affiliation{Department of Physics, University of Wisconsin, Madison, Wisconsin 53706, USA}
\author{M.~Ahlers}
\affiliation{Department of Physics, University of Oxford, 1 Keble Road, Oxford OX1 3NP, UK}
\author{K.~Andeen}
\affiliation{Department of Physics, University of Wisconsin, Madison, Wisconsin 53706, USA}
\author{J.~Auffenberg}
\affiliation{Department of Physics, University of Wuppertal, D-42119 Wuppertal, Germany}
\author{X.~Bai}
\affiliation{Bartol Research Institute and Department of Physics and Astronomy, University of Delaware, Newark, Delaware 19716, USA}
\author{M.~Baker}
\affiliation{Department of Physics, University of Wisconsin, Madison, Wisconsin 53706, USA}
% ...
\collaboration{IceCube Collaboration}\noaffiliation

\date{\today}

\begin{abstract}
A measurement of the atmospheric muon neutrino energy spectrum from 100 GeV to 400 TeV was performed using a data sample of about 18,000 up-going atmospheric muon neutrino events in IceCube.  Boosted decision trees were used for event selection to reject mis-reconstructed atmospheric muons and obtain a sample of up-going muon neutrino events.  Background contamination in the final event sample is less than $1\%$.  This is the first measurement of atmospheric neutrinos up to 400 TeV, and is fundamental to understanding the impact of this neutrino background on astrophysical neutrino observations with IceCube.  The measured spectrum is consistent with predictions for the atmospheric $\nu_ \mu + \bar \nu_ \mu$ flux.
\end{abstract}


\maketitle


\section{Introduction}

The IceCube neutrino telescope \cite{karle}, currently under construction in the glacial ice at the South Pole, is capable of detecting high energy neutrinos of all three flavors.  In particular, charged current (CC) interactions between $\nu_ \mu$ or $\bar \nu_ \mu$, and nucleons in the ice, produce muons.  IceCube detects the Cherenkov radiation produced as these muons propagate and undergo radiative losses.  By reconstructing the muon's track and energy loss, the direction and energy of the incident neutrino can be inferred.

Atmospheric neutrinos are produced in the decay chains of particles created by the interaction of cosmic rays with the Earth's atmosphere \cite{barr,honda,gaisser1}.  IceCube has an unprecedented high statistics, high energy reach for these atmospheric neutrinos.  Hence, IceCube can be used to test predictions for the flux of atmospheric neutrinos at high energies, including the uncertain contribution from charm production above about 100 TeV.  The atmospheric neutrino flux can also be used to verify that the IceCube detector is performing as expected \cite{francis}.  Understanding the energy and zenith dependence of the atmospheric neutrino flux in IceCube is important since this is an irreducible background for searches for a diffuse flux, or for point sources, of astrophysical neutrinos.

This analysis used data taken from April 2008, to May 2009, while IceCube was operating in a 40-string configuration.  The signal events were up-going atmospheric $\nu _\mu$ and $\bar \nu _\mu$ interactions.  The background was down-going atmospheric muons that were mis-reconstructed as up-going.  An as-yet unmeasured but anticipated diffuse flux of astrophysical neutrinos was ignored.  Predictions for this flux are negligible compared to predictions for the atmospheric neutrino flux, over most of the energy range for this analysis, and it can readily be accommodated within the reported uncertainties.

Boosted decision trees (BDT) were used to obtain an event sample with negligible background contamination.  An unfolding of the atmospheric neutrino energy spectrum, over the neutrino energy range 100 GeV to 400 TeV, was performed.  Systematic uncertainties in the unfolded spectrum were estimated and highlight the efforts that are underway to reduce systematic uncertainties in neutrino measurements with IceCube.

We will briefly review the production and distribution of atmospheric neutrinos in Section~\ref{atmos}.  In Section~\ref{icecube_det}, we will discuss the IceCube detector, and the detection of muon neutrino events in IceCube.  Event reconstruction and event selection specific to this analysis will be discussed in Section~\ref{data}.  The unfolding analysis and systematic uncertainties will be discussed in Section~\ref{unfolding_result}.  Finally, in Section~\ref{conclusion}, we will discuss the implications of the unfolded result.

\section{Atmospheric Neutrinos}
\label{atmos}
Cosmic rays are high energy particles, mostly protons and helium nuclei, but also heavier ionized nuclei, that are believed to be accelerated in various astrophysical phenomena \cite{pdg,gaisser}.  Possible cosmic ray production sites include active galactic nuclei, gamma ray bursts, and supernova explosions.  Detecting astrophysical neutrinos, produced in conjunction with cosmic rays at point sources such as these, is one of the primary goals of IceCube.  The energy spectrum of cosmic rays is rather steep, ${{dN} \mathord{\left/
 {\vphantom {{dN} {dE}}} \right.
 \kern-\nulldelimiterspace} {dE}} \propto E^{ - 2.7}$, and steepens to ${{dN} \mathord{\left/
 {\vphantom {{dN} {dE}}} \right.
 \kern-\nulldelimiterspace} {dE}} \propto E^{ - 3}$ above the ``knee'', or about $10^6$ GeV \cite{gaisser}.  A possible second knee is a steepening to about $E^{-3.2}$ above $5 \times 10^{8}$ GeV \cite{hoerandel}.  A further kink in the spectrum has been observed at $\sim 3 \times 10^{9}$~GeV, where the spectrum flattens to $dN/dE\propto E^{-2.7}$ again.  The event sample for this analysis is primarily the result of interactions of cosmic rays with energies below the first knee.

Hadronic interactions between cosmic rays and particles in the Earth's atmosphere produce large numbers of mesons, primarily pions and kaons.  Hundreds or even thousands of these mesons can be produced in the shower that follows the interaction of a single high energy cosmic ray.  Neutrinos are produced in the leptonic or semi-leptonic decays of charged pions or kaons, as well as in the subsequent decay of the muons.  Neutrinos from muon decay are important up to a few GeV.  Pions and kaons that decay in-flight are the primary source of atmospheric muon neutrinos from a few GeV up to about 100 TeV.  With rest-frame lifetimes on the order of $10^{-8}$~s, these mesons often lose some of their energy in collisions prior to decaying, leading to lower energy neutrinos among the decay products.  Hence, the spectral slope of this ``conventional'' atmospheric neutrino flux \cite{barr,honda} asymptotically becomes one power steeper than that of the primary cosmic ray spectrum.  Theoretical uncertainties in predictions for the conventional flux are dominated by uncertainties in the normalization and spectral distribution of the cosmic ray flux.  Additional uncertainties include the ratio of pions to kaons produced by cosmic ray interactions, which affects the zenith angle distribution, particularly near the horizon.

\begin{figure}
\includegraphics[width=3.5in]{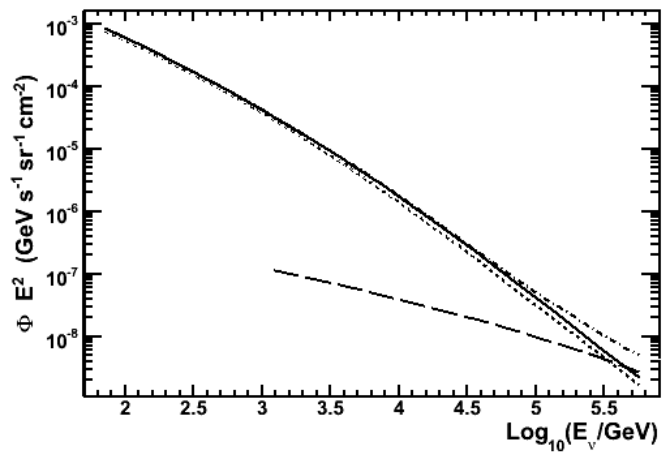}
\caption{The predicted flux of atmospheric muon neutrinos.  The solid line is the conventional $\nu_ \mu + \bar \nu_ \mu$ flux \cite{honda}, averaged over the zenith range $90^ \circ$ to $180^ \circ$.  The long-dashed line is the prompt $\nu_ \mu + \bar \nu_ \mu$ flux \cite{sarcevic}, also averaged over the zenith range $90^ \circ$ to $180^ \circ$.  The dot-dashed curve is the sum of the conventional and prompt models.  The flux predictions from \cite{honda} were extended to higher energies as discussed in Sect.~\ref{sim}.  For reasons discussed in Sect.~\ref{systematics}, the zenith region from $90^ \circ$ to $97^ \circ$ was not used in the analysis.  The zenith-averaged conventional flux, for the range $97^ \circ$ to $180^ \circ$, is included in the figure as the small-dashed line.  The prediction for the zenith-averaged prompt flux is not affected by this change in angular region.}
\label{gaisf25}
\end{figure}

At sufficiently high energies, another production mechanism is possible.  The ``prompt'' atmospheric neutrino flux \cite{sarcevic,martin,naumov} is made up of neutrinos produced in the semi-leptonic decays of charmed mesons and baryons.  These particles decay almost immediately (rest-frame lifetimes on the order of $10^{-12}$ s), before losing energy in collisions. Hence, the spectrum for the prompt flux more closely follows the cosmic ray spectrum and is about one power harder than the conventional flux at high energy.  The prompt flux has not yet been measured, but is expected to be important above about 100 TeV \cite{gaisser2,sarcevic}.  Just like the conventional flux, predictions for the prompt flux are impacted by uncertainties in the normalization and spectral distribution of the cosmic ray flux.  Additional sources of uncertainty for the prompt flux include charm production cross sections \cite{costa} and fragmentation functions, which have not been measured at these energies in accelerator experiments.  Figure~\ref{gaisf25} shows the predicted flux of conventional and prompt atmospheric muon neutrinos \cite{honda,sarcevic}.

Although high energy cosmic rays arrive almost isotropically, with deviations less than $0.1\%$ \cite{anisotropy}, the zenith angle dependence of high energy atmospheric neutrino production is complicated by the direction of the shower through the atmosphere.  The energy spectrum of nearly horizontal conventional atmospheric neutrinos is flatter than that of almost vertical neutrinos because pions and kaons in inclined showers spend more time in the tenuous atmosphere where they are more likely to decay before losing energy in collisions.   Additionally, attenuation of the neutrino flux by the Earth is a function of energy and zenith angle.  Above about 10 TeV, attenuation of the neutrino flux in the Earth is important, and affects the zenith and energy dependence of the flux at the detector.

\section{Neutrino Detection with IceCube}
\label{icecube_det}
\subsection{The IceCube Detector}
IceCube \cite{karle,str21} is able to detect neutrinos over a wide energy range, from about $100$ GeV to more than $10^{9}$ GeV.  The design is a balance between energy resolution, angular resolution, energy range, and cost, and was driven by the goal of detecting astrophysical neutrino point sources, which are believed to be correlated with cosmic ray production sites.  A large detector is required as a result of the extremely small cross-sections for neutrino interactions, as well as the low fluxes expected for astrophysical neutrinos.

When completed in 2011, IceCube will comprise 86 strings, with 5160 photomultiplier tubes (PMT).  Each string includes sixty digital optical modules (DOM).  A DOM is a single PMT and associated electronics in a glass pressure sphere.  The instrumented part of the array extends from 1450 m to 2450 m below the surface of the ice.  Horizontally, 78 of the strings are 125 m apart and spread out in a triangular grid over a square kilometer, so that the entire instrumented volume will be 1 ${\rm{km}}^3$ of ice.  Vertical DOM spacing is a uniform 17 m for these 78 strings.  A subset of the detector, known as ``DeepCore'', consists of eight specialized and closely spaced strings of sensors located around the center IceCube string.

\begin{figure*}
\includegraphics[width=5.0in]{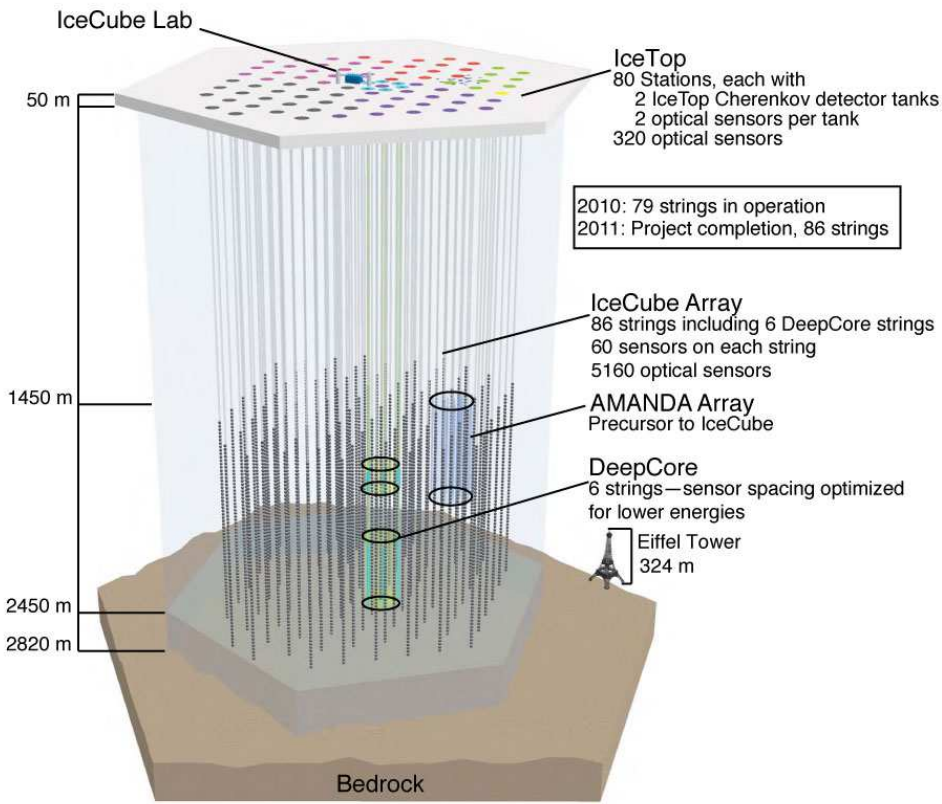}
\caption{IceCube Neutrino Observatory and its component arrays.}
\label{icecube}
\end{figure*}

\begin{figure}
\includegraphics[width=3.3in]{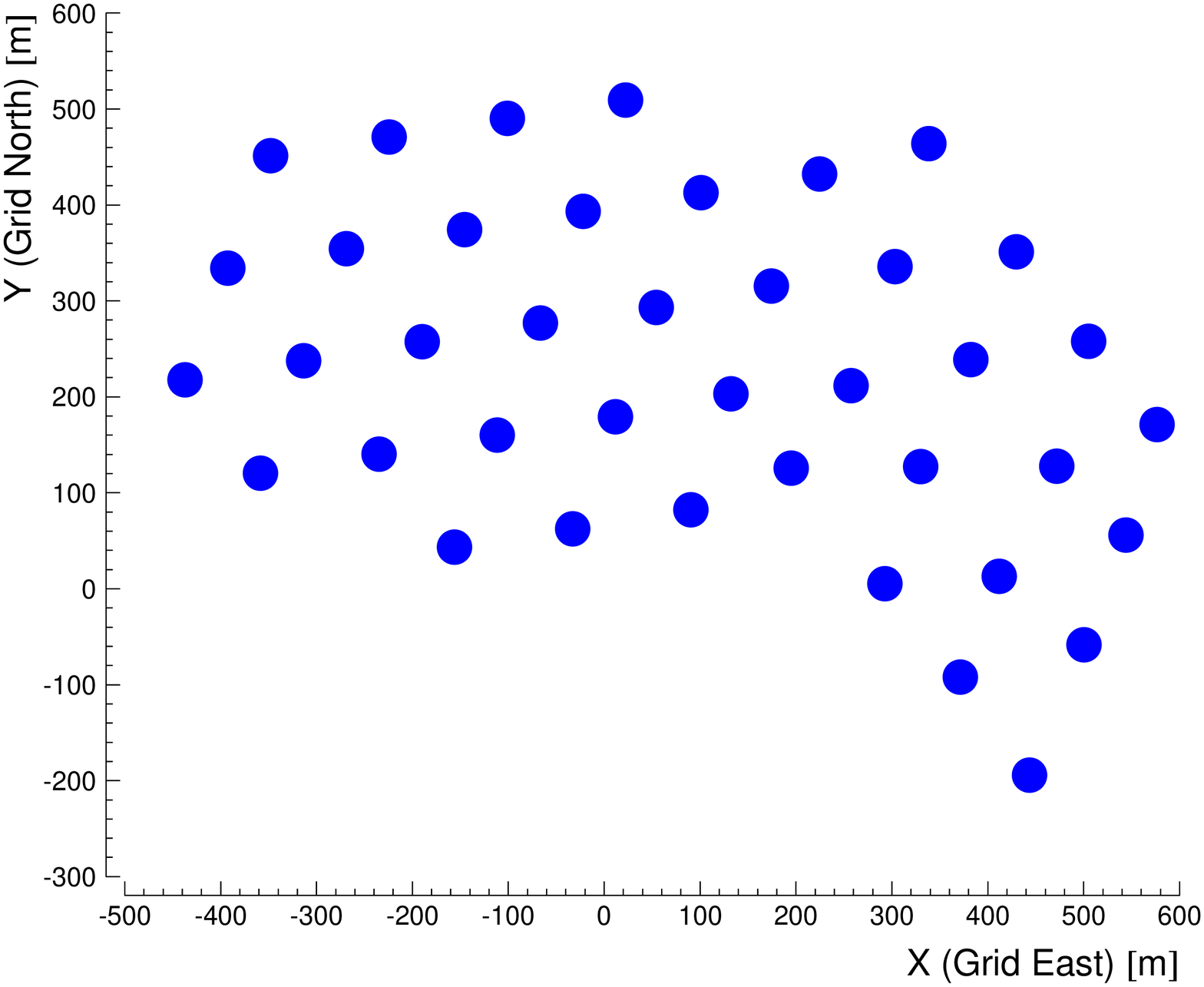}
\caption{Overhead view of IceCube 40 string configuration.}
\label{40_str}
\end{figure}

Figure~\ref{icecube} shows the IceCube observatory and its component arrays.  This analysis used data from 359 days of livetime while operating in a 40-string configuration, from April 2008, to May 2009.  Figure~\ref{40_str} shows an overhead view of the layout of the 40-string configuration, which was roughly twice as long in one horizontal direction as in the other.

\begin{figure}
\includegraphics[width=3.0in]{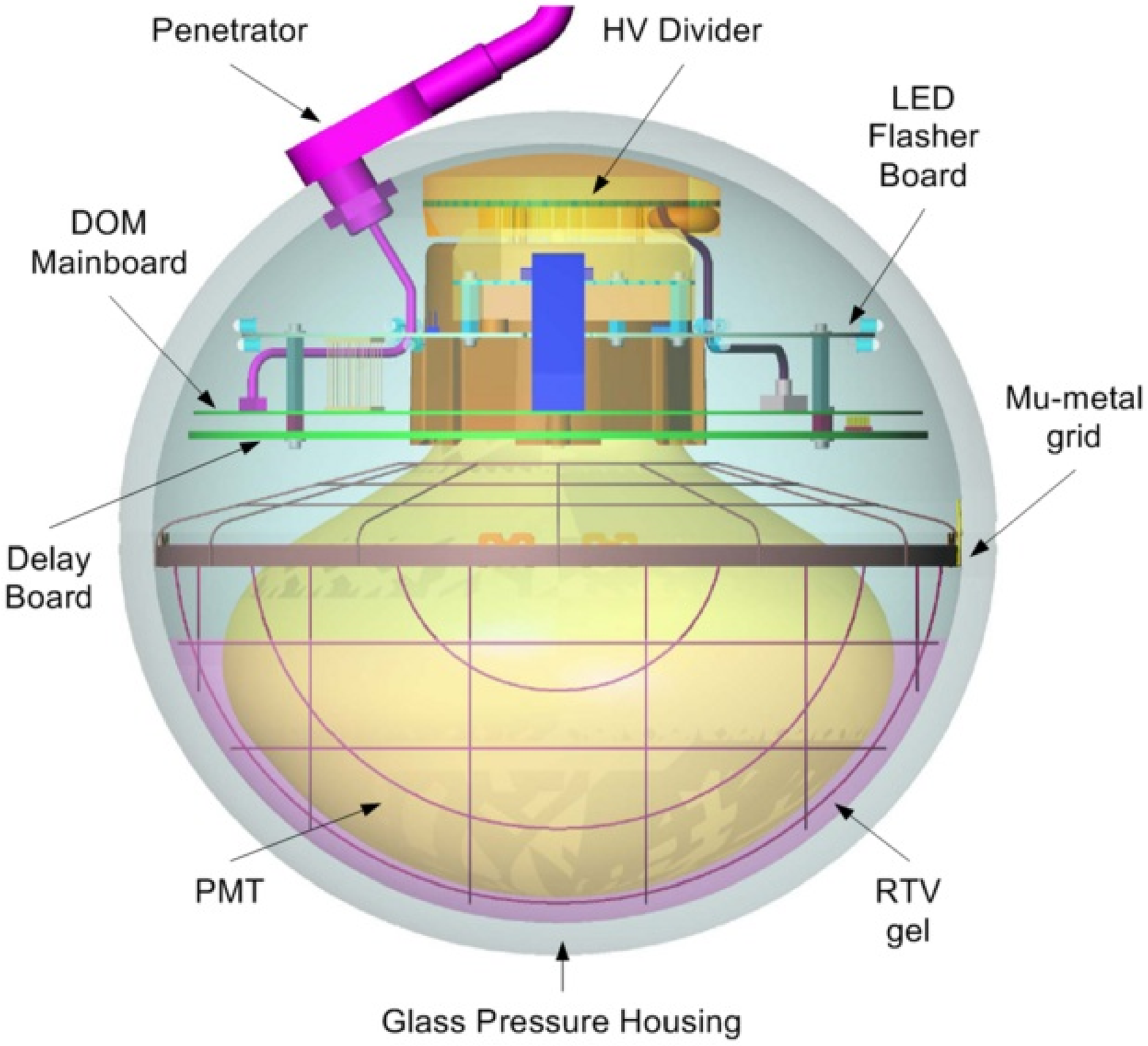}
\caption{Digital Optical Module.}
\label{dom_photo}
\end{figure}

At the heart of each DOM is a 10 inch (25 cm) Hamamatsu PMT \cite{pmt_paper} (see Fig.~\ref{dom_photo}).  A single Cherenkov photon arriving at a DOM and producing a photoelectron is defined as a hit.  DOM main board electronics \cite{daq} apply a threshold trigger to the PMT analog output.  This threshold is equivalent to 0.25 of the signal generated by a photoelectron, after amplification by the PMT.  When this threshold is exceeded, local coincidence checks between this DOM and nearest neighbor or next-to-nearest neighbor DOMs on a string are performed to reduce false triggers that result from dark noise.  If a nearest or next-to-nearest neighbor DOM also has a detection above threshold within a $\pm 1000$ ns window, the PMT total charge waveforms are digitized, time stamped, and sent to the surface.  The digitized waveform from a DOM can contain several pulses, and each pulse can be the result of multiple photoelectrons.  The simple majority trigger for building an event is eight hit DOMs within a 5000 ns trigger window.

The data rate from the data acquisition system at the South Pole far exceeds the amount of data that can be transmitted via satellite.  Hence, a significant reduction in the trigger-level data must be accomplished with software-based filtering at the South Pole.  A cluster of processors performs a variety of fast reconstructions on the data, and applies multiple software-based filters to the results.  These filters either reject events that are uninteresting background events, or extract particular classes of events.  Events are sent to a buffer if they pass one or more of the filters.  The transfer of data from this buffer over a communications satellite is handled by the South Pole Archival and Data Exchange (SPADE) system.

The deep glacial ice at the South Pole is optically transparent, making it an ideal medium for a large volume Cherenkov detector.  The ice sheet is just over 2800 m thick and was created over a period of roughly 165,000 years \cite{price}.  It serves multiple roles: a stable platform for the DOMs, the target medium for neutrino interactions, the propagation and detection medium for Cherenkov photons produced by charged particles, and an overburden for attenuation of down-going atmospheric muons.  Upward moving particles will have had to result from particles that penetrated the Earth and can readily be identified as resulting from neutrino interactions.

Optical properties of the ice are discussed in \cite{ice,price2,askebjer}.  Scattering and absorption of photons in the ice is caused by bubbles, dust particles, and crystal defects.  Below about 1400 m, the ice is essentially free of bubbles, and scattering is dominated by dust.  Micron-sized dust grains were carried as wind-borne aerosols during the periods of ice formation, and deposited in the ice.  Variations with depth are due to the periodic build up of dust that resulted from the prevailing atmospheric conditions when the layers of ice were being formed.

The depth and wavelength dependence of scattering and absorption as measured in the ice around the AMANDA detector is discussed in Ref.~\cite{ice}.  Now surrounded by IceCube and no longer operating, AMANDA was the predecessor and prototype for IceCube.  Ice properties were extrapolated to lower depths using ice core measurements taken at Vostok Station and Dome Fuji in Antarctica, then scaled to the location of IceCube using an age vs.\ depth relationship \cite{price}.  Studies are on-going that use LEDs on flasher boards within each DOM to directly measure ice properties in the deepest ice instrumented by IceCube.

\subsection{Muon Neutrino Detection}
\label{det}
Muon neutrinos undergoing CC interactions in the ice produce muons.  The muons on average carry about $75\%$ of the initial neutrino energy \cite{beacom}.  Simulation studies indicate that muon angular resolution is typically between $0.5^\circ $ and $1^\circ $, depending on the angle of incidence and the muon energy.  The energy loss per meter, for a muon propagating through the ice, is related to its energy \cite{pdg}:
\begin{equation}
 - \left\langle {\frac{{dE}}{{dX}}} \right\rangle  = \alpha(E)  + \beta(E) E,
\label{dedx_eqn}
\end{equation}
where $E$ is the muon energy, $\alpha  \approx 0.24$ GeV/m is the ionization energy loss per unit propagation length, and $\beta \approx 3.3 \times 10^{-4}$ ${\rm{m}}^{ - 1}$ is the radiative energy loss through bremsstrahlung, pair production, and photonuclear scattering, ($\alpha$ and $\beta$ are both weak functions of energy).  For muon energies less than about a TeV, energy loss is dominated by ionization, and the light produced is nearly independent of energy.  However, for higher energy muons, there are many stochastic interactions along the muon's path and there is a linear relationship between the energy loss per meter and the muon energy.  Most of the Cherenkov light emitted along the muon's path comes from the secondary particles produced in radiative losses.  An estimation of ${{dE} \mathord{\left/
 {\vphantom {{dE} {dX}}} \right.
 \kern-\nulldelimiterspace} {dX}}$, based on the amount of detected light, the event geometry, and the ice properties, was used in the energy spectrum unfolding discussed in Sect.~\ref{unfolding_result}.  The energy of individual events was not estimated.  Rather, the distribution of neutrino energies was directly inferred from the distribution of reconstructed muon $dE/dX$ values.

The detection rate for high energy $\nu _\mu$ ($\bar \nu _\mu$) is aided by the fact that the CC interaction cross section, as well as the range of the resultant muon, are proportional to the neutrino energy.  High energy muons have a significant path-length and can reach the detector even if produced outside of the detector, hence increasing the effective volume.  Muons in earth or ice can have a track length from several tens of meters, up to several kilometers, depending on the muon energy and the detection threshold.  The average track length, before the muon energy falls below a detection threshold ${E_\mu ^{\rm{th}} }$, is given by:
\begin{equation}
x_\mu   = \frac{1}{\beta }\ln \left[ {\frac{{\alpha  + \beta E_\mu  }}{{\alpha  + \beta E_\mu ^{\rm{th}} }}} \right],
\end{equation}
where $E _{\mu}$ is the initial muon energy.

\subsection{Simulation}
\label{sim}
Simulation of atmospheric muons and neutrinos was used for determining event selection and background rejection cuts.  Simulation was also used for the response matrix (discussed in Section~\ref{unfolding_result}) and the predicted $dE/dX$ distribution for the unfolding analysis.  Several specialized simulated data sets were used for systematics studies and toy Monte Carlo (MC) studies.

Muons from air showers were simulated with CORSIKA \cite{corsika}.  The primary cosmic ray energy spectrum known as the H{\"o}randel poly-gonato model \cite{hoerandel} was used.  In this model, the spectrum of each component is a combination of two power laws, with the turnover between the two power laws being a function of the nuclear charge $Z$ of the primary cosmic ray.  CORSIKA propagates cosmic ray primaries (up to Fe) to their point of interaction with a nucleus in the atmosphere.  Hadronic interactions in the atmosphere were modelled using the interaction model SIBYLL \cite{sibyll}.  Secondary particles were then tracked until they interacted or decayed. Coincident muons in the detector, originating from separate cosmic ray events, were accounted for by combining simulated events and re-weighting them to account for the probability of coincident events occurring.

Muon propagation and energy loss within and around the detector was simulated with the program MMC (Muon Monte Carlo) \cite{mmc}.  MMC accounts for ionization, bremsstrahlung, photo-nuclear interactions, and pair production.  In addition to muon tracks and energies, secondary particles from the stochastic energy losses are included in the output of MMC.  The production and propagation of Cherenkov light from the muons and secondary particles was simulated using the program Photonics \cite{photonics}, which accounts for the depth-dependent scattering and absorption properties of the ice.  Direct tracking of Cherenkov photons through the layered glacial ice was too computationally intensive for simulation production.  Photonics was run beforehand to create lookup tables which were then used during the detector simulation.  The tables included light yield and photon propagation time distributions at a given location in the ice from a given source type and location.  Simulation of the detector response to electromagnetic and hadronic showers (so-called cascade events) also used pre-tabulated light yield tables and photon propagation time information generated by Photonics.  An energy-dependent scaling factor was applied for hadronic cascades, to account for the fact that hadronic cascades produce less Cherenkov light than their electromagnetic counterparts \cite{cmc}.

Neutrino propagation from point of origin in the atmosphere to interaction in or near the detector was simulated with ANIS \cite{anis}.  ANIS generates neutrinos of any flavor according to a specified flux, propagates them through the Earth, and in a final step simulates neutrino interactions within a specified volume.  All simulated neutrinos were forced to interact, but their probability of interacting was included in the event weight assigned by ANIS.  ANIS accounts for CC and neutral current (NC) neutrino-nucleon interactions, as well as neutrino regeneration following NC interactions.  Also accounted for is the offset between neutrino propagation direction and the direction of the outgoing muon following a CC interaction.  Cross sections for $\nu_ \mu$ and $\bar \nu_ \mu$ CC and NC interactions were based on the CTEQ5 parton distributions \cite{cteq}.  The density profile in the Earth was modeled using the Preliminary Reference Earth Model \cite{earth}.

Simulated neutrino events were generated with an $E^{-2}$ spectral index, then weighted according to their contribution to the atmospheric neutrino flux.  The flux predictions of Honda {\it et al.} \cite{honda} were used for conventional atmospheric muon neutrinos, and those of Enberg {\it et al.} \cite{sarcevic} for prompt atmospheric muon neutrinos.  The predictions for muon neutrinos from pions and kaons were extended to higher energies by fitting a physics-motivated analytical equation based on energy and zenith angle (Ref.~\cite{gaisser2} and chapter 7 of Ref.~\cite{gaisser}) in an overlapping region with the detailed calculations of Honda {\it et al.} \cite{honda}.

\begin{figure}
\includegraphics[width=3.3in]{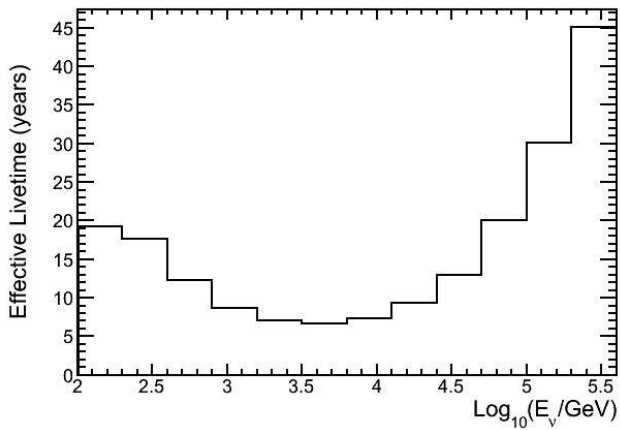}
\caption{Effective livetime of the simulation used to train the unfolding algorithm, as a function of neutrino energy.}
\label{eff_livetime}
\end{figure}

Since simulated events were generated with a harder spectrum than atmospheric neutrinos, the effective livetime for high energy events was boosted.  Additionally, since all events were forced to interact in or near the detector, the effective livetime for low energy events was boosted.  The effective livetime of the neutrino simulation used to train the unfolding algorithm is shown in Fig.~\ref{eff_livetime}.

\section{Event Reconstruction and Background Rejection}
\label{data}
A variety of algorithms are used in IceCube for event reconstruction, classification, and background rejection, depending on the energy range, the anticipated signal and backgrounds for a particular analysis, as well as the neutrino flavor.  The background for this analysis was down-going atmospheric muons that were mis-reconstructed as up-going.  Despite the depth of IceCube, the ratio of down-going atmospheric muons to muons produced in or near the detector by neutrino interactions is roughly one million to one \cite{karle}.  Below is a brief summary of the muon track reconstruction algorithms and event selection methods that were used in this analysis.  

\subsection{Event Reconstruction}
\label{reconstruction}
The LineFit reconstruction is a fast, first-guess algorithm based on the assumption that the Cherenkov photons from a muon propagate on a wavefront perpendicular to the track.  This assumption leads to a fitting algorithm that is extremely fast, and often estimates the muon track direction within ten degrees.  LineFit, and likelihood-based reconstructions (discussed next) seeded with the LineFit track, were used as part of the software filtering at the South Pole.  Additionally, the wavefront velocity estimated by LineFit is correlated with how well the track hypothesis fits the distribution of recorded light and was used as an event selection cut prior to one of the two BDTs.

Maximum likelihood reconstruction algorithms account for the geometric dependence of photon arrival times, as well as the stochastic variability in arrival times due to scattering in the ice.  The likelihood function to be maximized is the function \cite{ahrens2}
\begin{equation}
\mathcal{L} = \prod\limits_j {p\left( {{\bf{a}},t_{\rm{hit},j} } \right)} ,
\end{equation}
where ${\bf{a}}$ is the set of parameters characterizing the hypothesized track, i.e.\ three coordinates for the vertex location, two angles for the direction, and possibly energy, and ${p\left( {{\bf{a}},t_{\rm{hit}} } \right)}$ is the probability distribution function \cite{g_pandel} for photon hit times, given the track hypothesis.  The product is over all photon hits in the event.  In practice, the maximum of the likelihood function is found by minimizing the negative of the log of the likelihood, so the product becomes a sum.  To further simplify implementation, a transformation is made and time residual, $t_{\rm{res}} $, is used in place of hit time, $t_{\rm{hit}} $, where
\begin{equation}
t_{\rm{res}}  \equiv t_{\rm{hit}}  - t_{\rm{geo}}.
\end{equation}
The geometric travel time, $t_{\rm{geo}}$, is based on a straight photon path with no scattering.

Single Photoelectron (SPE) fits are likelihood reconstructions that use only the arrival time of the first photoelectron in all hit DOMs.  Typically, 16 or 32 iterations of the SPE fit are performed, with the seed track randomly altered for each iteration.  This helps ensure that a local minimum is not chosen as the final track.  The Multiple Photoelectron (MPE) fit is similar to the SPE fit, however, it uses the total number of observed photons to describe the arrival time of the first photon.  When many photons arrive at the same DOM, the first photon is scattered less than an average photon.  Since more information is used, the directional accuracy of the fit is often improved slightly, as compared to the SPE fit.  Moreover, using track quality parameters based on the MPE fit rather than on the SPE fit provided better event discrimination and improved the signal efficiency of the BDTs by about $10\%$.

\begin{figure}
\includegraphics[width=3.3in]{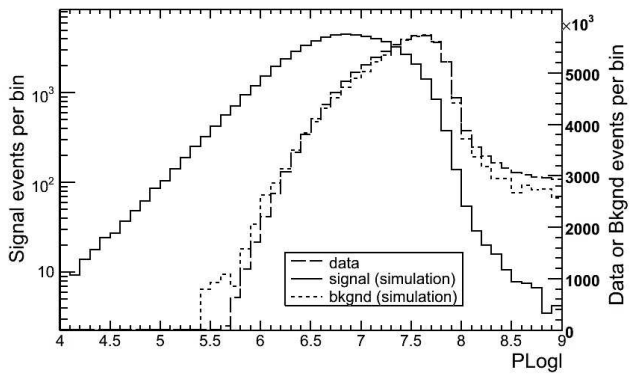}
\caption{Distribution of the PLogL variable (from the MPE fit), for neutrino simulation, muon background simulation, and for data.  A cut at a value of 8 based on the PLogL from a 32-iteration SPE fit has already been applied to reduce the amount of data requiring higher level processing.  PLogL was then recalculated based on the result of the MPE fit.}
\label{plogl}
\end{figure}

In addition to track location and direction, the likelihood reconstructions return several variables that are used to estimate fit quality.  These variables include the log-likelihood (LogL) and the reduced log-likelihood (RLogL).  ${\rm{RLogL}} = {\rm{LogL}}/n_{\rm{dof}} $, where ${n_{\rm{dof}} }$ is the number of degrees of freedom in the minimization, i.e.\ the number of hit DOMs minus the number of parameters to be fit.  RLogL is then (ideally) independent of the number of hit DOMs.  A similar scaled parameter called PLogL, equal to LogL/(number of hit DOMs - 2.5), has also been found to provide additional discriminatory power.  Figure~\ref{plogl} shows the distribution of the PLogL variable from the MPE fit, that was used by the BDTs.

\begin{figure}
\includegraphics[width=3.3in]{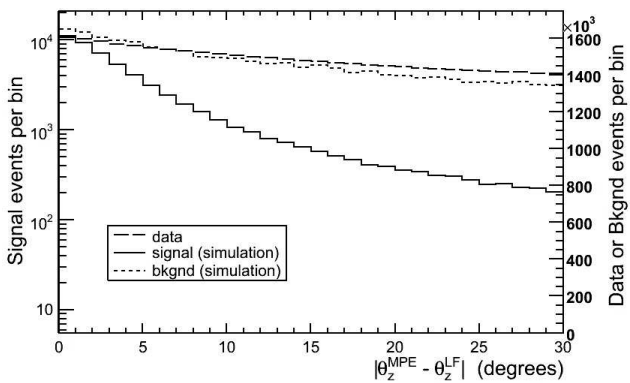}
\caption{Distribution of the difference between LineFit zenith angle and MPE fit zenith angle, for neutrino simulation, muon background simulation, and for data.}
\label{mpe_lf}
\end{figure}

Figure~\ref{mpe_lf} shows the distribution of the difference between LineFit zenith angle and MPE fit zenith angle, for signal simulation and for data.  This angular difference was used as an input to the BDTs.  

PhotoRec is a reconstruction algorithm that accounts for spatially variable ice properties \cite{grullon}.  It does this by incorporating light propagation tables created by Photonics \cite{photonics}.  The output of PhotoRec used in this analysis was the estimation of $dE/dX$, the average muon energy loss per unit propagation length (GeV/m) that would produce the detected amount of light. The reconstructed $dE/dX$ is proportional to the number of photons detected and hence to the number of photons emitted along the muon's track.  To correctly scale the proportionality, changes in the photon intensity due to the distance between the track and the hit DOMs, and the amount of scattering and absorption between light generation and detection points, are accounted for.  The reconstruction algorithm incorporates the detailed ice model, but assumes that stochastic energy losses are uniform along the track.  As mentioned in Section~\ref{det}, the amount of light emitted along the track of a high energy muon (greater than about a TeV) is linearly correlated with the muon's energy and it is possible to estimate the muon's energy (near the center of the detected track), using $dE/dX$, with an accuracy of 0.3 on a log scale.  However, for low energy muons, the amount of light emitted along the track is nearly independent of energy, as discussed in Section~\ref{det}.  Additionally, the PhotoRec algorithm does not account for the fraction of detected photons that may be from the hadronic shower at the interaction point (if that occurs inside the detector), nor does it account for the length of the muon track inside the detector.  Hence, the correlation between $dE/dX$ and muon energy degrades below a TeV.

The paraboloid algorithm \cite{parab} analyzes the value of the likelihood function around a seed track.  After transforming the coordinate space to one centered on the direction of the seed track, it fits a constant likelihood ellipse to the likelihood space around the direction of the track.  The important result is the paraboloid sigma, calculated from the major and minor axes of the constant likelihood ellipse. Paraboloid sigma provides an estimate of the pointing error of the track.

\begin{figure}
\includegraphics[width=3.3in]{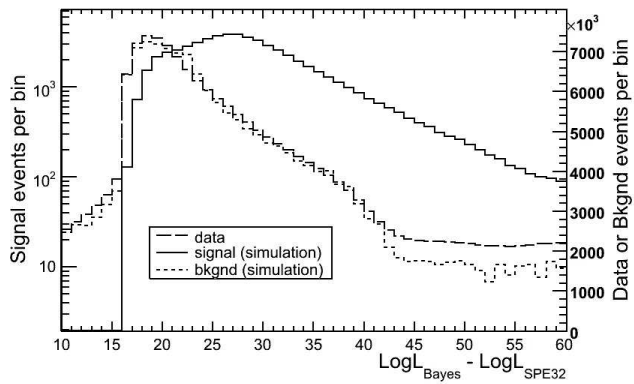}
\caption{Distribution of the Bayesian likelihood ratio, for neutrino simulation, muon background simulation, and for data.}
\label{bayes}
\end{figure}

In a Bayesian reconstruction, the standard likelihood function is multiplied by a bias function which depends only on the event hypothesis and not on the actual event data.  The bias is used as a way to include prior knowledge of the characteristics of the data, that mis-reconstructed down-going tracks dominate the signal by about three orders of magnitude at this stage.  The Bayesian likelihood ratio is the useful result from this reconstruction, ${\rm{LogL}}_{\rm{Bayes}}  - {\rm{LogL}}_{\rm{SPE32}}$, where SPE32 refers to a 32-iteration SPE fit.  Figure~\ref{bayes} shows the distribution of the Bayesian likelihood ratio, for simulation and for data.

In a reconstruction algorithm known as the umbrella fit, the minimizer is constrained to track directions with space angles more than $90^\circ$ from a seed track.  The likelihood ratio ${\rm{LogL}}_{\rm{Umbr}}  - {\rm{LogL}}_{\rm{SPE32}}$ is used as an event selection parameter.  Good tracks have a higher SPE likelihood than a fit constrained to have a directional component in the opposite direction.  This reconstruction provides discriminating power for certain events that are stuck in a local minimum in the likelihood space, such as down-going or near horizontal events that reconstruct as directly up-going.

\begin{figure*}
\includegraphics[width=6.5in]{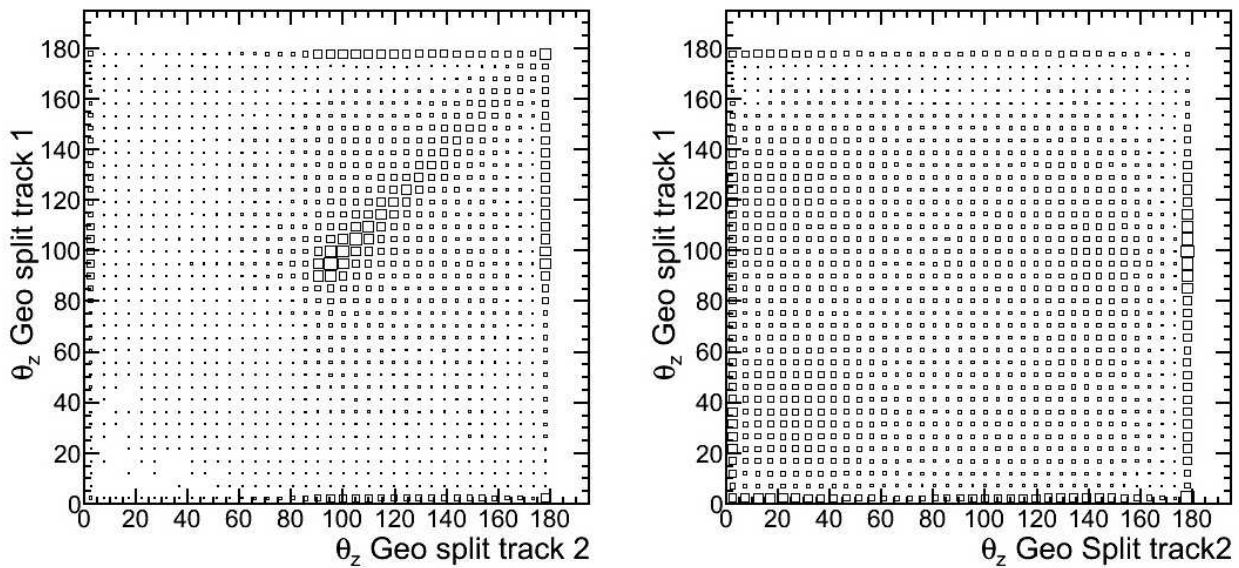}
\caption{Zenith angles from the SPE fits for the two sub-events created by the geometric split.  Neutrino simulation (left) and data (right).  Box size is proportional to the event density.}
\label{splits}
\end{figure*}

Split track reconstructions begin by creating four sub-events from the initial event.  Two sub-events are created by separating all hit DOMs into the group hit before the average time, and the group hit after the average time.  Two additional sub-events are based on geometry.  All hit DOMs are projected perpendicularly along the track.  Then, the DOMs are split into two groups based on whether they fall before or after the location of the center of gravity of the pulses.  LineFit and SPE reconstructions are performed on each of these four subsets.  These fits provide discrimination for poorly reconstructed tracks, as well as for tracks that reconstruct as up-going due to the superposition of hits from two separate down-going muons.  A loose cut on zenith angles from the split track reconstructions was used as an event selection cut prior to one of the two BDTs.  Additionally, the zenith angles were used as input variables for the BDTs.  Figure~\ref{splits} compares the zenith angles from the SPE fits for the two sub-events found by the geometric split.

In addition to zenith angles and likelihood ratios, several other measured or reconstructed variables were used for event discrimination.  For example, the likelihood that a track is properly reconstructed is correlated with the number of hit strings (NString), the more hit strings the better.

Photons originating from farther away from the DOM are more likely to have been scattered, and their associated distributions of arrival time probabilities are more spread out.  A larger number of direct hits, that is hits that propagate directly to the DOM with little or no scattering, has been found to be correlated with better track reconstruction.  The number of direct hits (NDir) is defined as the number of DOMs that have a hit with a residual time difference of $ - 15{\rm{ ns}} < t_{\rm{res}}  < 75{\rm{ ns}}{\rm{.}} $
The ratio of direct hits to the total number of detected pulses (NDir/NPulses) in an event was also used as a cut parameter.

The length of the event, (LDir), is determined by projecting hit DOMs onto the reconstructed track and calculating the distance between the two endpoints of the projection.  Larger values indicate a more reliable reconstruction of track direction.  LDir is calculated using direct hits only.

Smoothness is a measure of how well the observed hit pattern is explained by the hypothesis of constant light emission along the reconstructed muon track.  High quality tracks have hits equally spaced along the track.  This parameter, called SmoothAll, is calculated using all hits.  

\subsection{Filtering and Event Selection}
At trigger level, mis-reconstructed atmospheric muon events in the zenith region $90^ \circ$ to $180^ \circ$ outnumbered atmospheric neutrinos by a factor of about $10^ 5$.  These mis-reconstructed tracks were either individual muon tracks or coincident atmospheric muons that mimicked a single up-going event.  

Although a variety of filters were deployed at the South Pole for the 2008--2009 physics run, events used for this analysis were only required to pass the muon filter.  The muon filter was the primary filter for rejecting down-going atmospheric muons, and retaining generic $\nu_ \mu$ events from near or below the horizon.  Simple and fast reconstructions were performed in real-time at the South Pole. These initial reconstructions were less accurate than ones performed later, during off-line data processing.  However, they could be accomplished within the time and CPU constraints at the South Pole while keeping up with the trigger rate.  Zenith angles from LineFit and single-iteration SPE likelihood fits, as well as the number of hit DOMs (NChannel) and the average number of pulses per DOM, were used as selection variables in the muon filter.  After muon filter event selection was applied, background was reduced to a factor of about $10^4$ times the neutrino event rate.

Higher level reconstructions included improved likelihood reconstructions for better angular resolution and background rejection, as well as reconstruction of additional parameters, such as energy estimation.  Fits to additional track hypotheses were also performed.  Some higher level reconstructions incorporated the detailed ice model.  Prior to higher-level off-line data processing, events that were uninteresting or unusable, and that clearly were not going to pass final event selection, were removed by applying loose cuts based on the results of an SPE fit: zenith angle $> 80^ \circ$, RLogL $< 12$, and PLogL $< 8$.  This reduced the amount of background to roughly a factor of $10 ^3$ relative to signal.

Final event selection was accomplished with BDTs that used multiple reconstructed and observed parameters as input.

\subsection{Boosted Decision Tree Event Selection}
\label{bdt}
The Toolkit for Multivariate Data Analysis (TMVA) with ROOT \cite{tmva} was used to implement BDT event classification.  BDTs outperform straight cuts because the decision trees are able to split the phase space into a large number of hypercubes, each of which is identified as either signal-like or background-like \cite{tmva}.  Additionally, BDTs often out-perform other multivariate techniques because either there are not enough training events available for the other classifiers, or the optimal configuration (e.g.\ how many hidden layers for a neural network, which variables to use, etc.) is not known and is difficult to determine \cite{tmva}.  Testing with several different multivariate algorithms within TMVA indicated that the best results for separating signal from background in this case could be achieved with BDTs.  

The nodes of a decision tree form what looks like an inverted tree.  At each node, the algorithm chose the particular cut variable and cut value that provided the best discrimination between signal and background for the events in that node.  Events were then split into additional nodes that made up the next layer of the tree, and the process repeated until a minimum number of events in a node was reached.  Variables were used multiple times in a tree, with different cut values each time.  The final nodes were classified as signal or background, depending on the classification of the majority of training events that ended up in each node.

Boosting was used to overcome problems associated with statistical fluctuations in the simulation used to train the BDTs.  200 trees were derived from the same training ensemble by re-weighting events.  After one tree was created, events that were mis-classified in that tree had their weights increased, and the next tree was created.  This next tree then chose different variables and cut values at each node as a result of the altered weights.  The final classifier used a weighted average of the individual decisions of all 200 trees.

Two BDTs were used; one having better efficiency at lower energies, the other having better efficiency at middle and higher energies.  Events were accepted if their classification score from either BDT exceeded an optimized threshold.  The function of the BDTs was to distinguish between poorly reconstructed background events, and signal events that included some that were well reconstructed and some that were poorly reconstructed.  By applying pre-selection cuts prior to training the BDTs, some of the poorly reconstructed events were removed from the signal event samples, and the overall performance of each BDT was improved.  For the low energy BDT (BDT 1), the pre-selection cut was based on LineFit velocity (LineFit velocity $> 0.2c$).  For the other BDT (BDT 2), the pre-selection cut was based on zenith angles from the split track fits (all four zenith angles $> 80 ^\circ$).  The same cuts were applied to the actual data as were applied to the simulated background and signal event samples used for BDT training and testing.

Muon neutrino simulation with an $E^{-1}$ spectrum was used for signal events in the BDT training.  Although the true signal spectrum is much steeper than this, testing indicated this spectrum for training produced a BDT that performed better for higher energy events, with no compromise in performance for low energy events.  Cosmic ray muon simulation from CORSIKA was used for background events.  Following training, the BDTs were tested using independent signal and background event simulation.  Neutrino simulation weighted to an atmospheric spectrum, as well as single, double, and triple-coincident muon events, weighted to the cosmic ray muon spectrum, were used for testing the BDTs.

Table~\ref{bdt_vars} lists the specific variables used in the BDTs.  The NString variable was only used by BDT 1.  One additional difference between the two BDTs was the source of the Split Track fits.   For BDT 1, which was optimized for lower energies, the LineFit reconstructions for each of the four split tracks (two split geometrically and two split in time) were used.  For BDT 2, if 16-iteration SPE fits were successful for the split tracks, then those results were used, otherwise the LineFit results were used.  SPE fit results were not available for events in which there were too few hit DOMs in one or more of the splits to perform a likelihood fit.

\begin{table}[h]
\caption{Reconstruction variables used in the BDTs.  $\theta _Z$ refers to the zenith angle.  See Sect.~\ref{reconstruction} for explanations.}
\begin{ruledtabular}
\begin{tabular}{c}
BDT Variables \\
\hline
Paraboloid Sigma for the MPE fit\\
RLogL from the MPE fit\\
PLogL from the MPE fit\\
NDir\\
LDir \\
SmoothAll \\
NDir/NPulses \\
${\rm{LogL}}_{\rm{Bayes}}  - {\rm{LogL}}_{\rm{SPE32}} $ \\
NString \\
$\left| {\theta _Z^{\rm{MPE}}  - \theta _Z^{\rm{LineFit}} } \right|$\\
${\rm{LogL}}_{\rm{Umbr}}  - {\rm{LogL}}_{\rm{SPE32}} $\\
$\theta _Z $ from each of the Split-Tracks\\
\end{tabular}
\end{ruledtabular}
\label{bdt_vars}
\end{table}

\begin{figure*}
\includegraphics[width=6.5in]{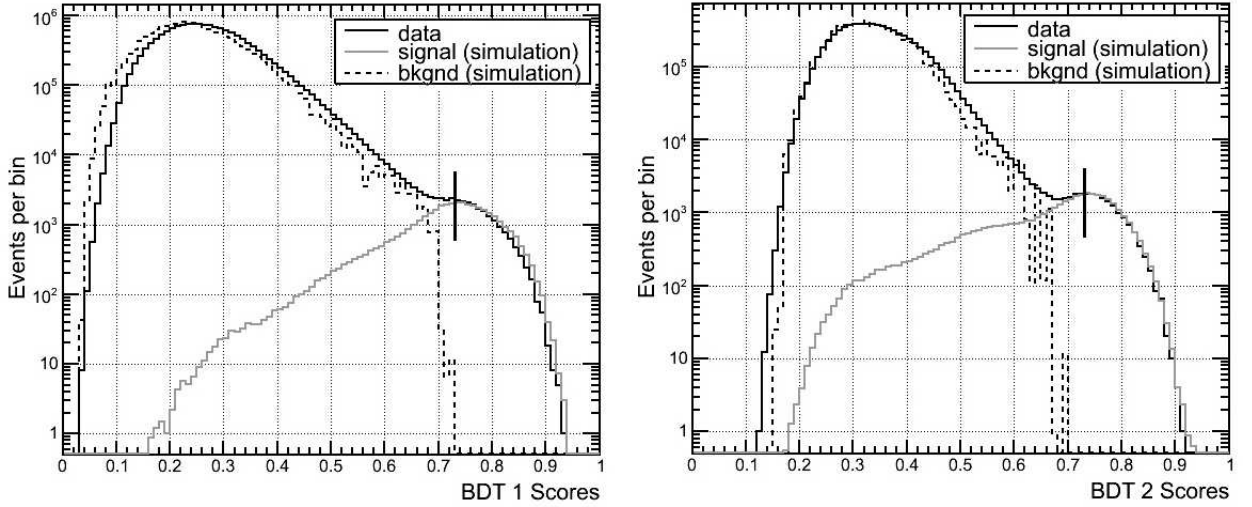}
\caption{Output of the BDTs for data, as well as neutrino and muon background simulation weighted to the same livetime (359 days).  The vertical lines mark the chosen cut value of 0.73 for each BDT.}
\label{bdt_scores}
\end{figure*}

Figure~\ref{bdt_scores} shows the output of each BDT, for the data and for simulation weighted to the same livetime (359 days).  The cut value of 0.73 was chosen to achieve greater than $99\%$ purity.  Testing the BDTs with simulated signal and background data sets indicated that the background contamination was less than $0.25\%$.  However, the effective livetime of the background simulation available for testing was not representative of a year of data.  The lack of sufficient background simulation near the chosen cut values can be seen in Fig.~\ref{bdt_scores}.  Because we did not have a reliable estimate of background contamination, comparisons between data and neutrino simulation were used to further verify that background rejection was performing as expected.  In particular, the data passing rate as a function of BDT cut values was compared to the predicted rate from atmospheric muon and neutrino simulation.  At looser BDT cut values, where sufficient simulated background events passed the BDT cuts to provide a statistically significant estimate, the background from simulation underestimated the apparent background in the data by about a factor of three.  Hence, the amount of background contamination in the final data set was conservatively estimated to be less than $1\%$.  The additional cut at a zenith angle of $97^ \circ$, discussed shortly, further reduced the potential for background contamination.

\begin{figure}
\includegraphics[width=3.3in]{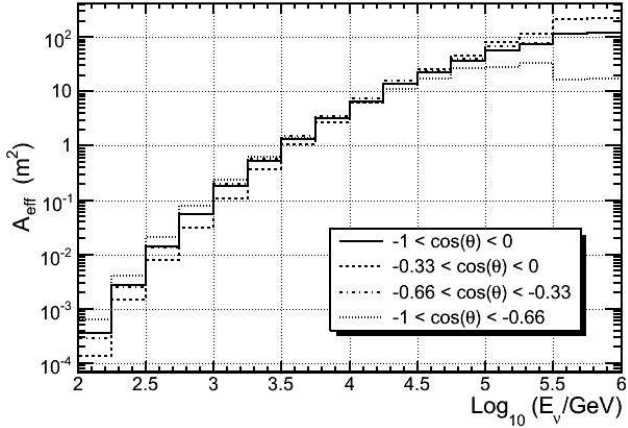}
\caption{Effective area for up-going muon neutrinos as a function of neutrino energy, for various zenith regions.}
\label{eff_area}
\end{figure}

\begin{figure}
\includegraphics[width=3.5in]{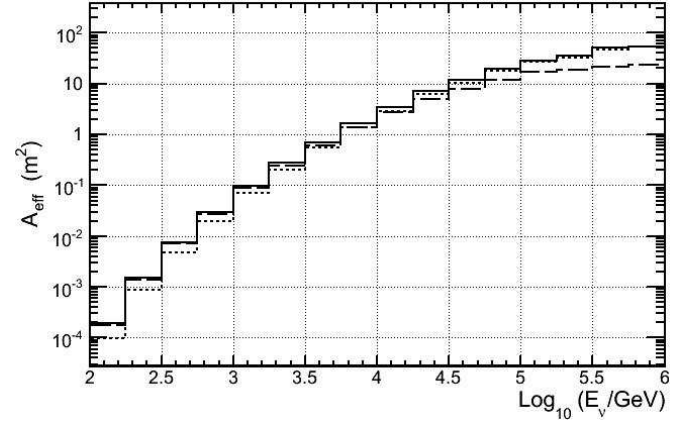}
\caption{Effective areas as a function of energy for each BDT.  BDT 1 (long-dashed line) performs better than BDT 2 at low energies while BDT 2 (short-dashed line) performs better than BDT 1 at higher energies.  Events are required to pass only one of the BDTs and the net effective area is the solid line.  In contrast to Fig.~\ref{eff_area}, this plot reflects the zenith-averaged effective area for the region $97^ \circ$ to $180^ \circ$.  This corresponds to the zenith region used for the analysis.}
\label{bdt_eff_area}
\end{figure}

The effective area, $A_{\rm{eff}}$, is the area occupied by a hypothetical detector with the same collecting power as IceCube, but with $100\%$ efficiency.  $A_{\rm{eff}}$ satisfies the equation
\begin{equation}
N_{events}  = \int {dt\int {d\Omega \int {dE \cdot \Phi (E,\theta )} } }  \cdot A_{\rm{eff}} (E,\theta ),
\end{equation}
where $N_{events}$ is the number of events passing final selection cuts and $\Phi (E,\theta )$ is the true flux of atmospheric neutrinos with units of ${\rm{GeV}}^{ - 1}\ {\rm{s}}^{ - 1}\ {\rm{ sr}}^{ - 1}\ {\rm{ cm}}^{ - 2}$.   In practice, the effective area is numerically calculated based on the number of neutrino events generated in simulation, the number passing final event selection cuts, and the event weights assigned in simulation to account for the probability of reaching and interacting in the detector.  Figure~\ref{eff_area} shows the effective area as a function of energy, for different zenith ranges, at the final cut level.  Figure~\ref{bdt_eff_area} shows the effective areas as a function of energy for BDT 1 and BDT 2 separately.

After eliminating data runs with some strings not operating, testing in progress, or various faults, there remained a total of 359 days of livetime from April 2008 to May 2009.  After final event selection cuts, the number of up-going neutrino events was 20,496, with zenith angles between $90^ \circ $ and $180 ^ \circ$.  An apparent excess of horizontal events in the data, or deficit in simulation, from $90^ \circ $ to $97^ \circ  $, will be discussed in more detail in Sect.~\ref{systematics}.  Since the origin of this mismatch could not be verified, an additional zenith angle cut was applied at $97^ \circ  $.  This resulted in a data sample of 17,682 atmospheric muon neutrino events from  $97^ \circ $ to $180^ \circ  $.

\section{Spectrum Unfolding}
\label{unfolding_result}
\subsection{Methodology}
\label{method}
The distribution of the energy-related observable, $dE/dX$, can be expressed as
\begin{equation}
b\left( {{{dE} \mathord{\left/
 {\vphantom {{dE} {dX}}} \right.
 \kern-\nulldelimiterspace} {dX}}} \right) = A\left( {E_\nu  ,{{dE} \mathord{\left/
 {\vphantom {{dE} {dX}}} \right.
 \kern-\nulldelimiterspace} {dX}}} \right)\Phi \left( {E_\nu  } \right),
\end{equation}
where $\Phi$ is a vector representing the true atmospheric neutrino flux as a function of energy, at the point of origin in the atmosphere, the vector $b$ is the distribution of $dE/dX$ for events in the final sample, and $A$ is the response matrix that accounts for the effects of propagation through the Earth, interaction in or near the detector, detector response, and event selection.  An analytical solution for $A$ is not known, so it is created from simulation.

The desired result from the energy spectrum unfolding is the true neutrino flux, $\Phi$.  Ideally, this could be determined by inverting the response matrix:
\begin{equation}
\label{resp_eqn}
\Phi \left( {E_\nu  } \right) = A^{ - 1} \left( {E_\nu  ,{{dE} \mathord{\left/
 {\vphantom {{dE} {dX}}} \right.
 \kern-\nulldelimiterspace} {dX}}} \right)b\left( {{{dE} \mathord{\left/
 {\vphantom {{dE} {dX}}} \right.
 \kern-\nulldelimiterspace} {dX}}} \right).
\end{equation}
However, direct solution is complicated by the fact that events are lost because the detector has limited efficiency (many neutrinos either do not interact near the detector or the events do not pass event selection cuts).  Additionally, the detector response is affected by limited energy resolution and there is significant smearing of events between bins (large off-diagonal elements in the response matrix).  Moreover, statistical fluctuations in the data can lead to unphysical variations in the unfolded spectrum.

The Singular Value Decomposition (SVD) unfolding algorithm \cite{hocker} was used to solve Eqn.~\ref{resp_eqn} and regularize the solution.  The SVD method involves factoring a non-invertible matrix into the product of two orthogonal matrices and a diagonal matrix, that can then be manipulated as necessary.  This algorithm has been implemented in the RooUnfold package \cite{roounfold} for use in the ROOT \cite{root} data analysis framework.  The inputs to the unfolding algorithm are the response matrix, the predicted histogram for the observed distribution, and a histogram for the expected true flux.  

\begin{figure}
\includegraphics[width=3.5in]{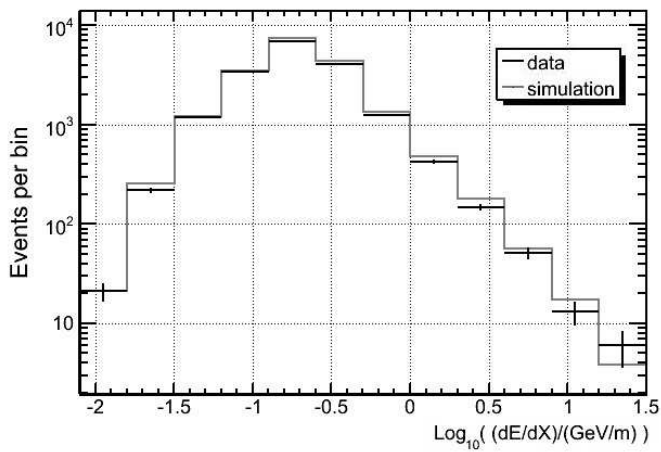}
\caption{Distributions of the $dE/dX$ observable, for data and for simulation.  The additional cut at zenith angle of $97^ \circ  $ has been applied.  Error bars for data are statistical only.}
\label{dedx_unf}
\end{figure}

The expected true flux, $\Phi_{_{MC} }$, is a 12 bin histogram binned in ${\rm{log}}_{10} (E_\nu/\rm{GeV})$ from 2 to 5.6, where $E_ \nu$ is the neutrino energy in GeV.  The predicted observables histogram, $b_{_{\rm{MC}} }$, is a 12 bin histogram of the expected $dE/dX$ distribution of events passing final cuts, binned in ${\rm{log}}_{10} ((dE/dX) / \rm{(GeV/m)})$ from $-2.1$ to $1.5$.  Figure~\ref{dedx_unf} shows the distributions, comparing data to simulation, for the observable $dE/dX$.  The response matrix, $A$, is a 12 by 12 histogram binned in ${\rm{log}}_{10} ((dE/dX) / \rm{(GeV/m)})$ vs.\ ${\rm{log}}_{10} (E_\nu/\rm{GeV})$, and filled with all events in $b_{_{MC} }$.  

The response matrix maps the distribution of reconstructed muon $dE/dX$ values to the distribution of neutrino energies.  The correlation between muon energy in the detector and the reconstructed $dE/dX$ is not exact, and neither is the correlation between muon energy in the detector and the incident neutrino energy.  The neutrino flux in or near the detector is affected by propagation through the Earth, during which CC interactions attenuate the neutrino flux and NC interactions alter the neutrino energy distribution.  The muon energy is only a fraction of the neutrino energy and only a fraction of the muon energy is observed.  Below about a TeV, where ionization rather than stochastic energy losses dominate, the energy loss rate is nearly independent of energy.  Additionally, the stochastic, radiative losses are not uniform along the muon's track, as assumed in the reconstruction algorithm.  If the muon is created in the detector, Cherenkov photons generated by the hadronic shower at the location of the CC interaction can be detected.  If the muon is created outside of the detector, it loses some of its energy before reaching the detector.  

\begin{figure}
\includegraphics[width=3.5in]{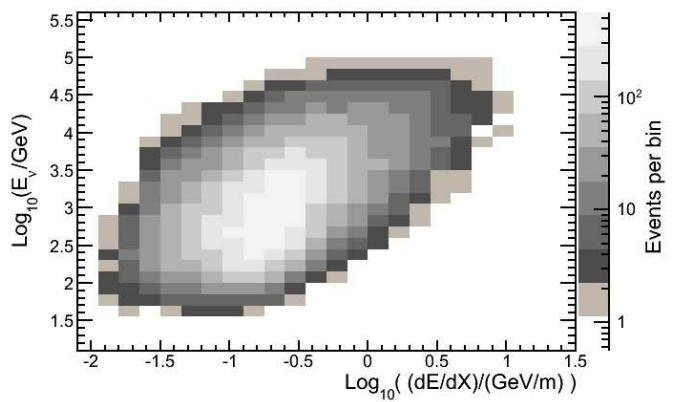}
\caption{Correlation between neutrino energy and the reconstructed muon $dE/dX$ observable, from simulation weighted to the atmospheric neutrino spectrum of \cite{honda,sarcevic}.}
\label{response}
\end{figure}

\begin{figure}
\includegraphics[width=3.5in]{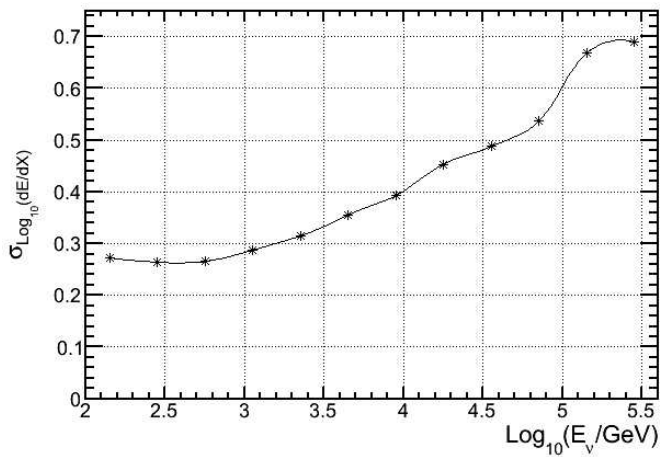}
\caption{Estimated neutrino energy resolution, from simulation.}
\label{resolution}
\end{figure}

Figure~\ref{response} shows the correlation between neutrino energy and reconstructed muon $dE/dX$.  An estimate of neutrino energy resolution, as a function of ${\rm{Log}}_{10} (E_\nu/\rm{GeV})$, is shown in Fig.~\ref{resolution}.  To estimate this resolution, a Gaussian fit was performed to the distribution of ${\rm{Log}}_{10} ((dE/dX) / \rm{(GeV/m)})$ in each of 12 ${\rm{Log}}_{10} (E_\nu/\rm{GeV})$ bins, and the standard deviations from these fits are shown in the figure.  At higher energies, the correlation between neutrino energy and reconstructed muon $dE/dX$ is hindered by the fact that the muon tracks are not contained within the detector and the muon can originate from a significant distance outside of the detector.  At lower energies, the resolution is aided by the fact that the events are more fully contained within the detector and the amount of detected light depends on the track length within the detector.

In addition to numerically inverting the response matrix, the unfolding algorithm applies smoothing (regularization) to the solution to ensure that statistical fluctuations in the data do not propagate as unphysical fluctuations in the unfolded spectrum.  The curvature in the solution, how sharply it can fluctuate from bin-to-bin, is regulated.  A regularization parameter enforces a smooth cutoff of higher frequency terms in the solution.  A lower cutoff biases the solution towards the shape of the expected spectrum, whereas a higher cutoff allows the solution to be influenced to a greater extent by fluctuations in the data.

The optimal choice of the regularization parameter depends on the number of bins and the sample size.  Two methods for determining the appropriate amount of regularization were used, as discussed in Ref.~\cite{hocker}.  The primary method used a result directly from the unfolding algorithm, where the coefficients of a particular decomposition of the rescaled measurement histogram were examined.  At lower indices these coefficients fall exponentially, and the critical term that determines the setting of the regularization parameter is at the end of the exponential fall, after which the coefficients are not significant.

As suggested in Ref.~\cite{hocker}, this result was checked using a series of toy simulations that were made systematically and statistically different from the expected true distribution.  The atmospheric neutrino flux models from Refs.~\cite{honda,sarcevic} were used as a baseline.  Variations in the spectral slope (up to $\pm$ 0.1) and normalization about this baseline were implemented.  For each underlying assumed true flux, many randomly fluctuated data sets were generated and each simulated data set was unfolded several times, using different choices for the regularization term.  The $\chi^2$ of each unfolded result relative to the true assumed spectrum was calculated and the distributions examined.  The regularization term giving the best average $\chi^2$ was the same as that found by the direct method using the decomposition coefficients.

\begin{figure}
\includegraphics[width=3.5in]{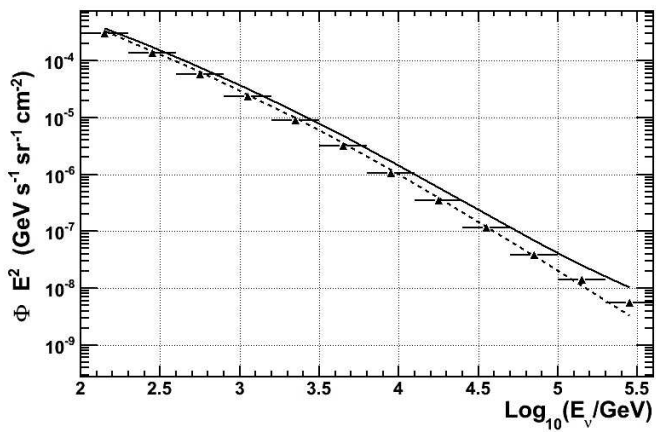}
\caption{Unfolding of known, toy spectrum.  The solid line is the assumed spectrum used for regularization, the sum of the conventional and prompt atmospheric neutrino flux models from Refs.~\cite{honda,sarcevic}.  The dashed line is the arbitrary, toy spectrum used for generating the toy data.  The unfolded result is shown without error bars.}
\label{toy_example}
\end{figure}

Fig.~\ref{toy_example} shows the performance of the unfolding algorithm to a toy spectrum.  In this example, a toy data set was created by arbitrarily modifying the event weights in simulation.  The spectral slope of the conventional atmospheric neutrino flux (Ref.~\cite{honda}) was made steeper by an index correction of -0.05, and the overall normalization was reduced to $80 \%$.  Additionally, the prompt flux was not included in the toy spectrum, creating a change in the shape of the energy spectrum at higher energies, where the shape of the actual flux is most uncertain.  As can be seen from Fig.~\ref{toy_example}, some bias is introduced by the regularization process at the highest energies where the event count is low and the shape of the true spectrum is different from the assumed spectrum. 

\subsection{Results and Systematic Uncertainties}
\label{systematics}
\begin{figure}
\includegraphics[width=3.5in]{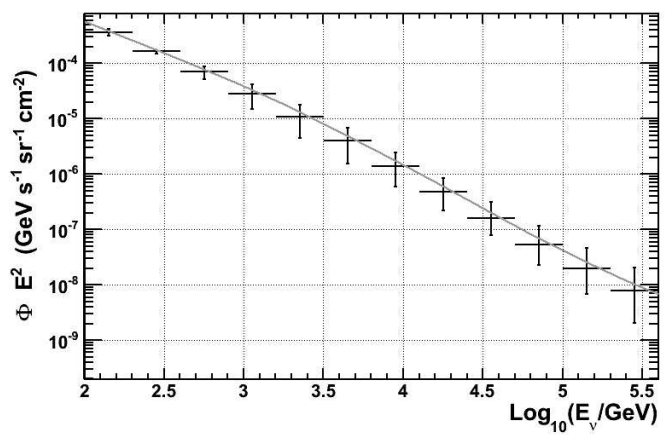}
\caption{Results of the atmospheric neutrino energy spectrum unfolding.  The unfolded spectrum is shown in black, vertical lines are the estimated uncertainties.  The gray line is the spectrum that provided the expected shape for the regularization, and includes the conventional atmospheric neutrino flux according to Honda {\it et al.} \cite{honda} and the prompt flux according to Enberg {\it et al.} \cite{sarcevic}.  This is the zenith-averaged $\nu_ \mu + \bar \nu_ \mu$ flux for the region $97^ \circ - 180^ \circ $. }
\label{unfolding}
\end{figure}

The results of the atmospheric neutrino energy spectrum unfolding can be seen in Fig.~\ref{unfolding}.  Event selection cuts that isolated track-like events, caused by the muons created in $\nu_ \mu$ CC interactions, eliminated localized events from the electromagnetic showers induced by $\nu _e$ CC interactions and the hadronic showers induced by NC interactions.   Additioanlly, production of $\nu_ \tau$ ($\bar \nu_ \tau$) by cosmic rays is negligible.  IceCube is not able to distinguish between neutrino and antineutrino events.  Hence, the unfolded spectrum is the sum of $\nu_ \mu$ and $\bar \nu_ \mu$, averaged over the zenith region $97^ \circ - 180^ \circ $.

The major uncertainties in the unfolded spectrum are from four categories.  These are uncertainties in DOM sensitivity and ice properties, zenith-dependent data/simulation inconsistencies, statistical uncertainties and the impact of the regularization process, and miscellaneous normalization errors such as neutrino cross section and muon energy loss uncertainties.  The bin-by-bin values for estimates of each of these error sources were added in quadrature to obtain the final uncertainty estimate for each bin of the unfolded flux.

Systematic uncertainties in ice properties and DOM sensitivities lead to systematic errors in the distribution of reconstructed $dE/dX$ values, as well as the energy dependence of the detector's effective area.  To estimate the impact of these uncertainties, two specialized neutrino simulation datasets were created.  In one dataset, the number of photons striking each DOM was boosted by $10 \%$.  In the other dataset, the number of photons was reduced by $10 \%$.  From this, it was found that a $\pm 10\%$ change in the photon flux leads to a $\pm 15\%$ change in event rate and a $\pm 0.09$ change in the ``apparent'' spectral index of the neutrino flux.  These factors were found from a three parameter fit that determined the changes in normalization, spectral index, and zenith angle tilt, of standard atmospheric neutrino simulation, to reproduce the best fits to the $dE/dX$ distributions of the simulated event samples from these specialized datasets.

The uncertainty in the DOM sensitivity is taken as $\pm 8\%$, based on the measured uncertainty in the PMT sensitivity \cite{pmt_paper}.  The $\pm 10\%$ change in photon flux in the specialized simulation is effectively the same as a $\pm 10\%$ change in PMT sensitivity, so the normalization and spectral index correction factors just mentioned were scaled to the $\pm 8\%$ for PMT sensitivity uncertainty.  To apply these values to ice property uncertainties, the change in the average number of photons striking a DOM that would result from a change in the ice properties had to be estimated.

First, we assumed a mean propagation length $L_{_p }  = 30$ m with $\pm 10\%$ uncertainty \cite{ice}.  The propagation length for diffusive photon transport from a point source is defined as
\begin{equation}
L_p  = \sqrt {\frac{{L_e L_a }}{3}} ,
\end{equation}
where $L_e$ is the effective scattering length and $L_a$ is the absorption length.  Then, we estimated the fractional change in the number of photons at a distance $d$ from the muon's track as
\begin{equation}
\frac{{N' - N}}{N} \sim e^{\frac{{ \pm .1d}}{{(1 \pm .1)L _p }}}  - 1, 
\end{equation}
where $ N\sim\left( {{1 \mathord{\left/ {\vphantom {1 d}} \right. \kern-\nulldelimiterspace} d}} \right)e^{{{ - d} \mathord{\left/ {\vphantom {{ - d} {L_p }}} \right. \kern-\nulldelimiterspace} {L_p }}} $ is the number of photons at distance $d$ for the nominal propagation length, and $ N'\sim\left( {{1 \mathord{\left/ {\vphantom {1 d}} \right. \kern-\nulldelimiterspace} d}} \right)e^{{{ - d} \mathord{\left/ {\vphantom {{ - d} {\left( {1 \pm .1} \right)L_p }}} \right. \kern-\nulldelimiterspace} {\left( {1 \pm .1} \right)L_p }}} $ is the number of photons at distance $d$ for the perturbed propagation length (nominal $\pm 10\%$).  The average distance between the track and the hit DOMs, per event, was estimated from simulation to be about 35~m.  The net result of this approximation was that the uncertainty in the average photon flux reaching the DOMs was estimated to be $\pm 12\%$ on average, as a result of ice property uncertainties.  The normalization and spectral index correction factors from the specialized simulated data sets were scaled to this $\pm 12\%$ uncertainty in the photon flux.

It should be pointed out that this method of estimating the impact of ice property uncertainties is affected by two approximations.  First, the accuracy of the diffuse flux equation is limited at ranges less than several propagation lengths.  Second, changes in ice properties would also change the distribution of photon arrival times at the hit DOMs, an effect which is not accounted for in the specialized simulation data sets.  Comparisons were made between these simulated data sets and simulation generated using Photonics tables derived from a modified ice model.  In this modified ice model, scattering and absorption in the cleaner layers of the ice were arbitrarily reduced.  This comparison, as well as preliminary results from ongoing work to improve the simulation of photon propagation in the ice and derive a more accurate estimate of uncertainties related to ice properties, indicated that the method used here likely over-estimates the impact of ice property uncertainties on the normalization and apparent spectral index, particularly in the higher energy bins.

\begin{figure}
\includegraphics[width=3.5in]{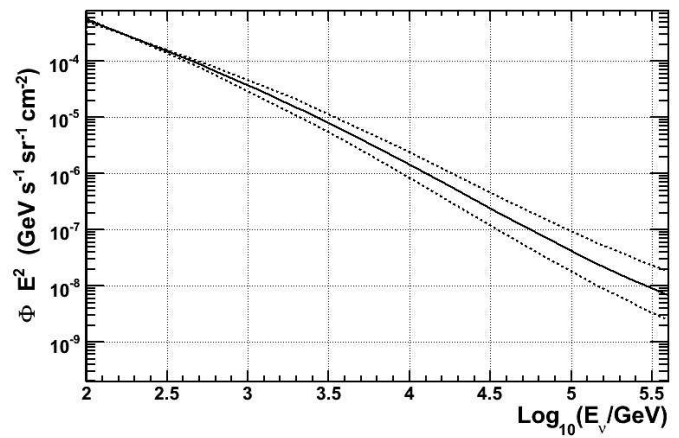}
\caption{Possible variability in the true neutrino flux consistent with DOM sensitivity and ice property uncertainties.  The solid line is the predicted atmospheric neutrino flux (\cite{honda,sarcevic}).  The dashed lines are the maximum and minimum of the possible range of variability consistent with DOM sensitivity and ice model uncertainties.  As mentioned in the text, work is on-going to reduce this range of uncertainty.}
\label{unf_omice}
\end{figure}

Adding the uncertainties in detector response due to ice properties and DOM sensitivity in quadrature leads to an estimated $\pm 22\%$ uncertainty in the normalization, correlated with an uncertainty of $\pm 0.13$ in the apparent spectral index.  These detector uncertainties lead to uncertainties in the apparent neutrino flux.  For a given detector response, i.e.\ a measurement of the $dE/dX$ distribution, the true normalization and spectral index of the neutrino flux cannot be constrained better than allowed by these uncertainties.  Figure~\ref{unf_omice} shows the resulting range of uncertainty in the measurement of the atmospheric neutrino energy spectrum.

\begin{figure}
\includegraphics[width=3.5in]{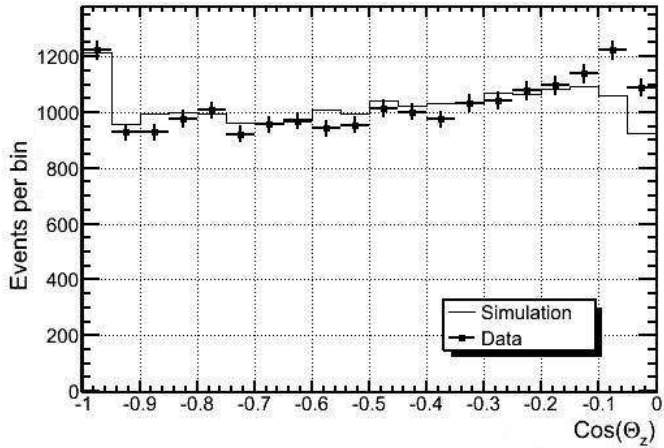}
\caption{Cosine($\theta_Z$) distributions for data and for simulation, using zenith angle from the MPE fit.  Simulation has been normalized to the data.  Error bars for data are statistical only.}
\label{cos_zen}
\end{figure}

As mentioned in Sect.~\ref{bdt}, there was a statistically significant excess of events in data, or a deficit in simulation, between $90^ \circ - 97^ \circ  $, i.e near the horizon.  Figure~\ref{cos_zen} shows the cos(zenith) distributions for data and for simulation, with simulation normalized to the data.  A similar excess was also observed in the AMANDA detector \cite{amanda_unf}.  A number of checks and tests were performed, including evaluation of track quality parameters, the depth-dependence of the excess, the strength of the BDT scores, and visual examination of a subset of events in a software-based event viewer.  The horizontal excess in data does not decrease with depth, nor with tightened BDT cuts.  If the BDT cut is loosened, mis-reconstructed muons show up predominantly near the top of the detector, as expected.  These checks are consistent with the possibility that the excess events are 
due to muons from atmospheric neutrino interactions.  However, it is also possible that they are due to an excess of mis-reconstructed atmospheric muons.

It is likely that the lower event rate in simulation, close to the horizon, is due to uncertainties in the simulation of Cherenkov photon propagation in the ice or of inaccuracies in the simulation of cosmic ray events, such as insufficient livetime, limitations with the cosmic ray model or its implementation in CORSIKA, or uncertainties in muon propagation and energy loss. Hence, it cannot be excluded that the horizontal excess is due to residual and un-simulated atmospheric muons and coincident events. It could also be related to uncertainties in the atmospheric neutrino flux due to atmospheric variability, discussed shortly. Since we were not able to verify the precise origin of the mismatch near the horizon, events in the zenith region $90^ \circ$  to $97^ \circ  $ were not used.

\begin{figure}
\includegraphics[width=3.5in]{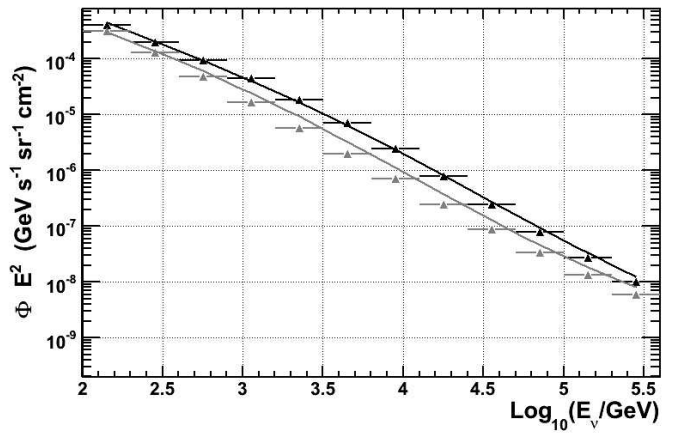}
\caption{Comparison of unfolded energy spectra for different zenith ranges.  Separate unfoldings were performed for the zenith range  $97^ \circ - 124^ \circ $ (black) and the zenith range  $124^ \circ - 180^ \circ  $ (gray).  The unfolded results for each region (horizontal lines) and the predicted spectrum corresponding to each region (curves) are shown.  Uncertainties for these results are not shown.}
\label{unf_anis}
\end{figure}

To estimate the impact of any remaining zenith-dependent systematic uncertainties in the zenith range $97^ \circ - 180^ \circ $, separate unfoldings were performed for the zenith range  $97^ \circ - 124^ \circ $ and the zenith range  $124^ \circ - 180^ \circ  $.  The results of this test are shown in Fig.~\ref{unf_anis}, together with the predicted zenith-averaged flux corresponding to each angular range.  The differences between result and prediction are not consistent between the two regions.  For the more vertical events (gray in Fig.~\ref{unf_anis}), the flux is lower than predicted for middle and higher energies.  For the more horizontal events (black in Fig.~\ref{unf_anis}), the flux is slightly lower than predicted at low and at high energies.  The relative differences between result and prediction for the two zenith regions was taken as an estimate of the impact of anisotropic uncertainties.

Seasonal and regional variations in the atmospheric temperature profile are expected to lead to variations in the atmospheric neutrino flux \cite{seasonal}, and could be causing the direction-dependent differences between data and simulation.  Colder temperatures correspond to a greater air density and a shallower atmosphere.  Greater atmospheric density leads to more collisions of pions and kaons prior to their decay.  Hence, the production of high energy neutrinos is reduced.  The converse occurs for warmer temperatures.  The kinematics of collision and decay, and slant angle through the atmosphere, conspire to lead to variations in the energy and zenith angle dependence of atmospheric neutrino production for different atmospheric conditions.  The normalization uncertainty on the Honda {\it et al.} conventional atmospheric neutrino flux model includes an estimated $3\%$ uncertainty due to uncertainties in the atmospheric density profile \cite{honda,sanuki}.  However, the flux calculation uses a climatological average atmosphere (the US-standard '76).  The estimate of the error in the flux calculation is based on the error in the climatological average atmospheric density profile.  It does not account for changes in normalization, or in the energy and zenith distribution of atmospheric neutrinos, that results from regional and seasonal atmospheric variability.

The impact of statistical uncertainties in the data, as well as bias due to the regularization process and the possibility that the assumed spectrum used to compute the amount of regularization may be different from the true spectrum, were estimated using toy simulations.  First, a six-parameter forward folding fit to the data was performed.  In the forward folding fit, the general form of the flux was assumed to be consistent with the shape of the theoretical predictions \cite{honda,sarcevic}, but corrections to the normalization, spectral index, and zenith angle tilt of the conventional and prompt atmospheric neutrino flux models were propagated through simulation.  The fit variables that produced the best fit between the simulated detector response and the data were used to re-weight simulated events in the toy experiments to mimic the data.  The results of the forward folding fit indicated a possible systematic suppression of the neutrino event rate at higher energies, and this suppression was included in the toy simulations.

\begin{figure}
\includegraphics[width=3.5in]{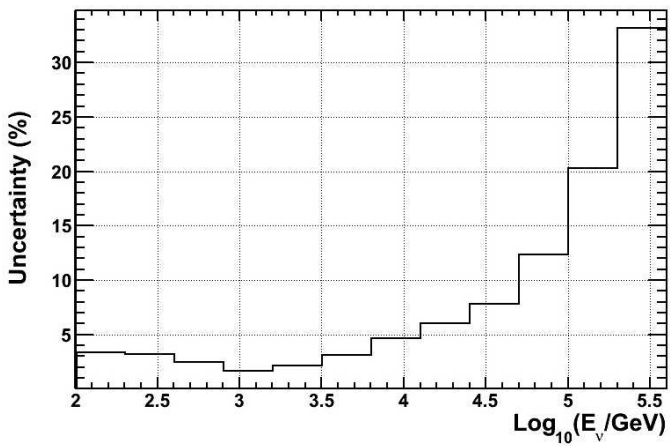}
\caption{Statistical and regularization-induced uncertainties in the unfolded result.  The errors in each bin are given as the percent of the true flux.}
\label{unf_unf}
\end{figure}

One thousand trials were performed, with events in each bin of the toy $dE/dX$ distributions fluctuated according to a Poisson distribution.  Statistical uncertainties in the neutrino simulation were also included.  The difference between the unfolded energy spectrum and the known, ``true'' spectrum that the toy experiments were based on were computed for each trial.  The $68^{th}$ percentile of the errors in each bin from the 1000 trials were assigned as the uncertainty.  The result of this analysis of statistical and regularization uncertainties is shown in Fig.~\ref{unf_unf}, where the errors in each bin are given as percent of the true flux.  A potential systematic bias between the shape of the true flux and the shape of the assumed flux used to train the unfolding algorithm accounts for roughly half of the uncertainty indicated in Fig.~\ref{unf_unf} for the two highest energy bins.

A $3\%$ uncertainty in the charged current, deep-inelastic neutrino-nucleon scattering cross section is estimated to lead to a $3\%$ uncertainty in atmospheric neutrino event rates, and uncertainties in muon energy loss are estimated to lead to a $1\%$ uncertainty \cite{five_year}.  Reconstruction and cut biases are estimated to introduce a $2\%$ uncertainty in event rate.  Adding these, and the $1\%$ background contamination, in quadrature gives a $4\%$ uncertainty in the event rate, assumed to be independent of energy.

\begin{figure}
\includegraphics[width=3.5in]{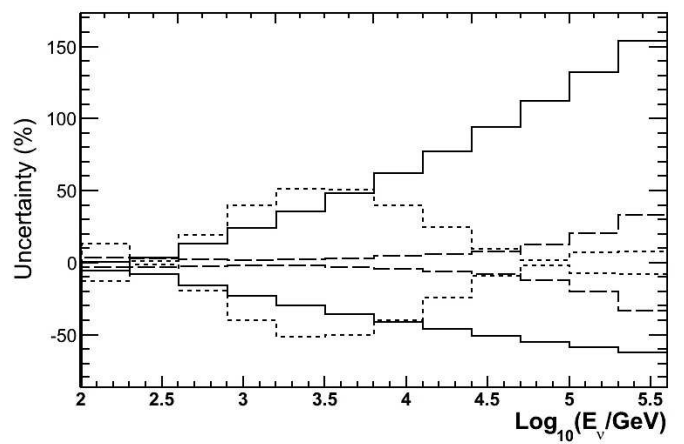}
\caption{Sources of uncertainty in the unfolded energy spectrum.  The solid lines are the systematic uncertainties due to DOM sensitivity and ice property uncertainties, the short-dashed lines are the uncertainties implied by zenith-dependent inconsistencies in data/simulation comparisons, and the long-dashed lines are the statistical and regularization uncertainties from toy MC studies.  Not shown is the uniform $4\%$ uncertainty due to miscellaneous normalization errors assumed to be independent of energy.}
\label{unf_errs_1}
\end{figure}

A summary of uncertainties in the unfolded result can be seen in Fig.~\ref{unf_errs_1}, as well as Table~\ref{summary}.  At the lower end of the unfolded energy range, uncertainties are dominated by zenith-dependent inconsistencies.  At the middle of the range, uncertainties are dominated by the DOM sensitivity and ice property uncertainties, as well as the zenith-dependent uncertainties.  Uncertainties in DOM sensitivity and ice properties dominate at higher energies.

\begin{table}[h]
\caption{Zenith-averaged, unfolded atmospheric muon neutrino energy spectrum.}
\begin{ruledtabular}
\begin{tabular}{c|c|c}
$\log _{10} \left( {{{E_\nu  } \mathord{\left/
 {\vphantom {{E_\nu  } {{\rm{GeV}}}}} \right.
 \kern-\nulldelimiterspace} {{\rm{GeV}}}}} \right)$
&$
\begin{array}{*{20}c}
   {{{dN} \mathord{\left/
 {\vphantom {{dN} {dE_\nu  }}} \right.
 \kern-\nulldelimiterspace} {dE_\nu  }} \cdot E_\nu ^2 }  \\
   {\left( {{\rm{GeV \, s}}^{ - 1} {\rm{ \, sr}}^{ - 1} {\rm{ \, cm}}^{ - 2} } \right)}  \\
\end{array}
$
& $\% $ Uncertainty \\
\hline
2.0 - 2.3 & $3.6 \times 10^{ - 4} $ & +29, -28\\
2.3 - 2.6 & $1.6 \times 10^{ - 4} $ & +21, -22\\
2.6 - 2.9 & $7.0 \times 10^{ - 5} $ & +31, -32\\
2.9 - 3.2 & $2.8 \times 10^{ - 5} $ & +50, -50\\
3.2 - 3.5 & $1.1 \times 10^{ - 5} $ & +65, -62\\
3.5 - 3.8 & $4.0 \times 10^{ - 6} $ & +71, -63\\
3.8 - 4.1 & $1.4 \times 10^{ - 6} $ & +74, -58\\
4.1 - 4.4 & $4.7 \times 10^{ - 7} $ & +82, -53\\
4.4 - 4.7 & $1.6 \times 10^{ - 7} $ & +95, -53\\
4.7 - 5.0 & $5.4 \times 10^{ - 8} $ & +113, -57\\
5.0 - 5.3 & $2.0 \times 10^{ - 8} $ & +135, -64\\
5.3 - 5.6 & $7.9 \times 10^{ - 9} $ & +158, -72\\
\end{tabular}
\end{ruledtabular}
\label{summary}
\end{table}

\begin{figure*}
\includegraphics[width=5.5in]{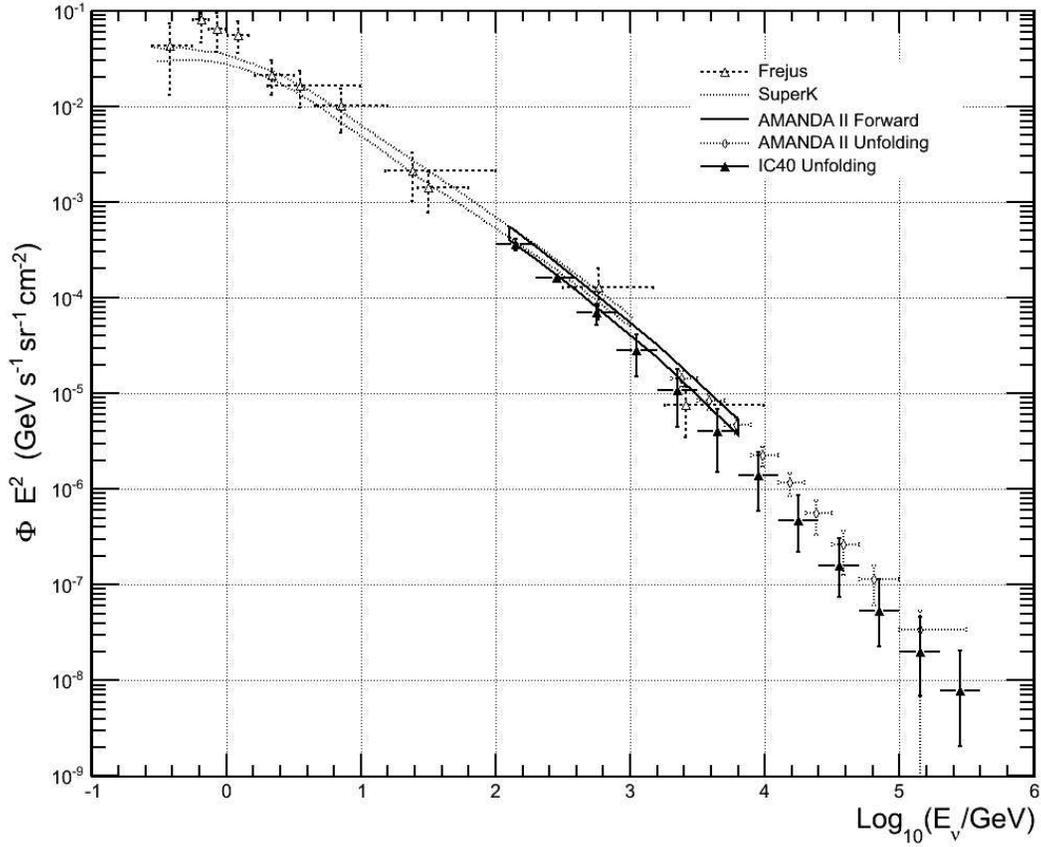}
\caption{Comparison with previous measurements of the atmospheric neutrino energy spectrum; the Fr\'{e}jus result \cite{frejus}, upper and lower bands from SuperK \cite{superk}, an AMANDA forward folding analysis \cite{amanda_fwd}, and an AMANDA unfolding analysis \cite{amanda_unf}.  All measurements include the sum of neutrinos and antineutrinos.  The AMANDA unfolding analysis was a measurement of the zenith-averaged flux from $100^ \circ$ to $180^ \circ  $.  The present analysis (IC40 unfolding), which extends the measurement up to 400 TeV, is a measurement of the zenith-averaged flux from $97^ \circ $ to $180^ \circ  $.   Vertical error bars include systematic as well as statistical uncertainty.}
\label{compare}
\end{figure*}

\begin{figure}
\includegraphics[width=3.5in]{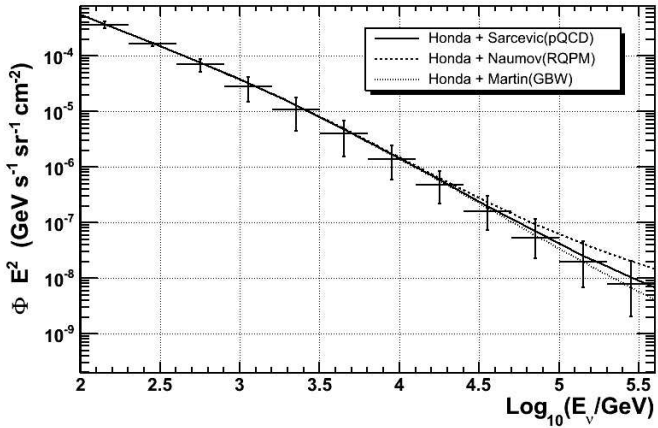}
\caption{Comparison of various prompt flux models to the unfolded result.  The models shown are the sum of the Honda flux \cite{honda}, plus one of Sarcevic \cite{sarcevic}, Naumov \cite{naumov}, or Martin \cite{martin}.  Vertical error bars include systematic as well as statistical uncertainty. }
\label{prompt}
\end{figure}


\section{Conclusions and Outlook}
\label{conclusion}
A zenith averaged unfolding of the atmospheric muon neutrino flux ($\nu _\mu$ plus $\bar \nu _\mu$), from 100 GeV to 400 TeV, was performed.  This is the first atmospheric neutrino measurement to such high energies, and the spectrum is consistent with predictions for the atmospheric muon neutrino flux.  However systematic uncertainties will need to be reduced before specific flux models \cite{barr,honda,sarcevic,martin,naumov} can be constrained.  In particular, we are as yet unable to confirm the contribution of a prompt flux.  Figure~\ref{compare} compares the results of this analysis (IC40 unfolding) to previous measurements of the atmospheric neutrino energy spectrum.  As discussed in Sect. \ref{systematics}, the estimate of uncertainties in the IceCube result are dominated by DOM sensitivity and ice property uncertainties, as well as the zenith-dependent mismath between data and simulation.  These uncertainty estimates are expected to be reduced as our simulation is improved.  A comparison between the unfolded spectrum and various prompt flux models \cite{sarcevic, naumov, martin} is shown in Fig.~\ref{prompt}.  

Several improvements are anticipated in atmospheric neutrino measurements with IceCube.  Correlations between variations in atmospheric temperature profiles, and the energy and zenith angle dependence of the atmospheric neutrino flux, are being investigated using in-situ atmospheric temperature measurements.  Pulsed LED sources installed on each DOM are being used to extend the ice description to the deepest ice in the detector with in-situ measurements like those done in AMANDA for the ice down to 2100 m.  Studies with cosmic ray muons are being used to reduce the uncertainty in DOM sensitivity.  Work is also on-going to identify and correct potential problems in simulation that could be contributing to data/simulation mismatch.  Perhaps most significantly, this includes improving the simulation of light propagation within the detector, which is anticipated to improve the data to simulation agreement for several measured and reconstructed variables.  These improvements will be discussed in a future paper.  Once simulation of light propagation in the ice has been improved, it should be possible to use a more sophisticated and realistic method for estimating the impact of ice model uncertainties.  As data collection continues, and improvements to simulation are implemented, it will be possible to extend the measurement of the atmospheric neutrino energy spectrum with IceCube to PeV energies, as well as significantly reduce the uncertainties.

\begin{acknowledgments}

We acknowledge the support from the following agencies:
U.S. National Science Foundation-Office of Polar Programs,
U.S. National Science Foundation-Physics Division,
University of Wisconsin Alumni Research Foundation,
the Grid Laboratory Of Wisconsin (GLOW) grid infrastructure at the University of Wisconsin - Madison, the Open Science Grid (OSG) grid infrastructure;
U.S. Department of Energy, and National Energy Research Scientific Computing Center,
the Louisiana Optical Network Initiative (LONI) grid computing resources;
National Science and Engineering Research Council of Canada;
Swedish Research Council,
Swedish Polar Research Secretariat,
Swedish National Infrastructure for Computing (SNIC),
and Knut and Alice Wallenberg Foundation, Sweden;
German Ministry for Education and Research (BMBF),
Deutsche Forschungsgemeinschaft (DFG),
Research Department of Plasmas with Complex Interactions (Bochum), Germany;
Fund for Scientific Research (FNRS-FWO),
FWO Odysseus programme,
Flanders Institute to encourage scientific and technological research in industry (IWT),
Belgian Federal Science Policy Office (Belspo);
University of Oxford, United Kingdom;
Marsden Fund, New Zealand;
Japan Society for Promotion of Science (JSPS);
the Swiss National Science Foundation (SNSF), Switzerland;
A.~Gro{\ss} acknowledges support by the EU Marie Curie OIF Program;
J.~P.~Rodrigues acknowledges support by the Capes Foundation, Ministry of Education of Brazil.

\end{acknowledgments}


\end{document}